\documentclass[hidelinks]{article}

\usepackage{arxiv}

\usepackage[utf8]{inputenc} % allow utf-8 input
\usepackage[T1]{fontenc}    % use 8-bit T1 fonts
\usepackage{hyperref}       % hyperlinks
\usepackage{url}            % simple URL typesetting
\usepackage{booktabs}       % professional-quality tables
\usepackage{amsfonts}       % blackboard math symbols
\usepackage{nicefrac}       % compact symbols for 1/2, etc.
\usepackage{microtype}      % microtypography
\usepackage{lipsum}		% Can be removed after putting your text content
\usepackage{graphicx}
\usepackage{natbib}
\usepackage{doi}
\usepackage{graphicx}
\usepackage{amsmath, amssymb}
\usepackage{color}
\usepackage[normalem]{ ulem }
\usepackage{soul}

\title{Bayesian generalized method of moments applied to pseudo-observations in survival analysis}

\author{ \href{https://orcid.org/0009-0007-7389-7612}{\includegraphics[scale=0.06]{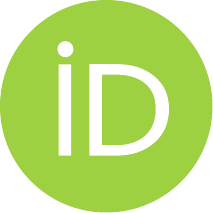}\hspace{1mm}Léa~Orsini} \\
	Oncostat U1018, Inserm\\
	University Paris-Saclay\\
	Villejuif, France \\
	\texttt{lea.orsini@gustaveroussy.fr} \\
	\And
	\href{https://orcid.org/0000-0002-0196-4157}{\includegraphics[scale=0.06]{orcid.pdf}\hspace{1mm}Caroline~Brard} \\
	Ipsen Innovation, Clinical Development Organisation\\
	Les Ulis, France \\
	\texttt{caroline.brard@ipsen.com} \\
	    \And
	    \href{https://orcid.org/0000-0002-3747-6905}{\includegraphics[scale=0.06]{orcid.pdf}\hspace{1mm}Emmanuel~Lesaffre} \\
	    I-Biostat\\
        KU-Leuven\\
	    Leuven, Belgium \\
	    \texttt{emmanuel.lesaffre@kuleuven.be} \\
	    \And
	    \href{https://orcid.org/0000-0003-3276-1392}{\includegraphics[scale=0.06]{orcid.pdf}\hspace{1mm}Guosheng~Yin} \\
	    Department of Statistics and Actuarial Science\\
        The University of Hong Kong\\
        Hong Kong, China\\
	    \texttt{gyin@hku.hk} \\
	    \And
	    \href{https://orcid.org/0000-0001-8960-5345}{\includegraphics[scale=0.06]{orcid.pdf}\hspace{1mm}David~Dejardin} \\
	    Product Development, Data Sciences\\
        F. Hoffmann-La Roche AG\\
	    Basel, Switzerland. \\
	    \texttt{david.dejardin@roche.com} \\
     \And
     	\href{https://orcid.org/0000-0003-3292-0939}{\includegraphics[scale=0.06]{orcid.pdf}\hspace{1mm}Gwénaël~Le Teuff} \\
        Oncostat U1018, Inserm\\
	    University Paris-Saclay\\
	    Villejuif, France \\
	    \texttt{gwenael.leteuff@gustaveroussy.fr} \\
}

\date{}

\renewcommand{\headeright}{}
\renewcommand{\undertitle}{}
\renewcommand{\shorttitle}{Bayesian generalized method of moments applied to pseudo-observations in survival analysis}

\hypersetup{
pdftitle={Bayesian generalized method of moments applied to pseudo-observations in survival analysis},
pdfauthor={Léa~Orsini, Caroline~Brard, Emmanuel~Lesaffre, Guosheng~Yin, David~Dejardin, Gwénaël~Le Teuff},
pdfkeywords={Bayesian analysis, Generalized method of moments, Pseudo-observations, Survival analysis},
}

\begin{document}
\maketitle

\begin{abstract}
Bayesian inference for survival regression modeling offers numerous advantages, especially for decision-making and external data borrowing, but demands the specification of the baseline hazard function, which may be a challenging task. We propose an alternative approach that does not need the specification of this function. Our approach combines pseudo-observations to convert censored data into longitudinal data with the Generalized Methods of Moments (GMM) to estimate the parameters of interest from the survival function directly. GMM may be viewed as an extension of the Generalized Estimating Equation (GEE) currently used for frequentist pseudo-observations analysis and can be extended to the Bayesian framework using a pseudo-likelihood function. We assessed the behavior of the frequentist and Bayesian GMM in the new context of analyzing pseudo-observations. We compared their performances to the Cox, GEE, and Bayesian piecewise exponential models through a simulation study of two-arm randomized clinical trials. Frequentist and Bayesian GMM gave valid inferences with similar performances compared to the three benchmark methods, except for small sample sizes and high censoring rates. For illustration, three post-hoc efficacy analyses were performed on randomized clinical trials involving patients with Ewing Sarcoma, producing results similar to those of the benchmark methods. Through a simple application of estimating hazard ratios, these findings confirm the effectiveness of this new Bayesian approach based on pseudo-observations and the generalized method of moments. This offers new insights on using pseudo-observations for Bayesian survival analysis.
\end{abstract}

% keywords can be removed

\newpage
\section{Introduction}
\label{s:intro}

Bayesian analysis offers many benefits in pharmaceutical research for drug development and clinical trials. The Bayesian computation methods are flexible and allow for fitting every model by estimating the posterior distribution of the parameters through a sampling procedure, even when no closed-form formula is available for this particular problem. Consequently, it allows the estimation of the posterior tail probability for any given threshold that may be clinically relevant, which is particularly useful for decision-making \citep{Held2020}. It also provides adaptive design methods for clinical trials that are naturally suited for interim analysis. In addition, Bayesian methods are advantageous in the context of rare diseases or precision medicine, where external information can be incorporated through the prior definition \citep{Lesaffre2024}.

Despite those advantages, Bayesian survival analysis is yet rarely used in survival analysis (\citet{Chevret2011}, \citet{Brard2017}). One reason may be that contrary to the frequentist framework where the partial likelihood of the Cox proportional hazard model can be used to estimate the regression coefficients of covariates on the survival outcome from right censored data \citep{Cox1972}, in Bayesian inference, the baseline hazard function is usually modeled and associated with priors \citep{Biard2021}, introducing nuisance parameters in this setting. Numerous Bayesian models have been proposed using parametric distributions (exponential or Weibull) and other functions (monotone or polynomials) referenced in \citet{Ibrahim2001}, Chapters 2 and 3. According to the literature review of \citet{Fors2020}, the most common model in randomized clinical trials is the piecewise exponential model, which assumes the baseline hazard function to be constant on intervals. More complex models have been developed using splines, which allow more flexibility \citep{Murray2016}. Non-parametric alternatives exist but involve many parameters and are computationally intensive. The Gamma process is chosen in many applications, and Cox's partial likelihood can be seen as the limiting case of this Bayesian process by allowing the prior precision to approach zero \citep{Kalbfleisch1978}. 

Over the past twenty years, the use of pseudo-observations in the frequentist framework has become an attractive research field in survival analysis since it offers a flexible and unique framework to directly estimate quantities of interest such as the survival probability, the cumulative incidence, the transition and state-occupation probabilities in multi-states models, or the restricted mean survival time \citep{Andersen2010}. Pseudo-observations are computed for a specific quantity of interest and a straightforward regression model, with pseudo-observations as outcome, is used to directly estimate the association between the covariates and this quantity. Although other approaches exist to model these different quantities, covariate adjustment is difficult with non-parametric estimators, while the assumption of fully parametric estimators may be challenged by the data \citep{Sachs2022}. Thus, by transforming (right or interval) censored data into pseudo-observations, survival analysis turns into a standard regression problem. 

When traditional Bayesian survival methods involve the formulation of the full likelihood, including the specification of the baseline hazard function in the setting of regression coefficient estimation, which may be challenging due to censoring, the transformation of censored data into pseudo-observations may be advantageous in overcoming this issue. This paper presents a methodology to analyze pseudo-observations in the Bayesian framework, creating an alternative approach for Bayesian survival analysis, which does not require specifying the baseline hazard function. Currently, pseudo-observations are usually analyzed as an outcome of a generalized linear model using the Generalized Estimating Equations (GEE), introduced in \citet{Liang1986}. This marginal approach does not involve a likelihood function and is consequently not easily translatable to the Bayesian framework. Our approach relies on the Generalized Methods of Moments (GMM) for which a Bayesian version based on a pseudo-likelihood function has been developed by \citet{Yin2009}. The GMM method has been defined by \citet{Hansen1982} and is widely used in econometrics. GMM is defined by specifying multiple moments from the data. GMM estimates are obtained by minimizing a quadratic inference function that combines these moments.

In this paper, we assess the usefulness of the frequentist and Bayesian GMM in the particular context of estimating hazard ratios with pseudo-observations. The rest of the paper is organized as follows. Section \ref{s:Methods} presents the theoretical aspects of pseudo-observations analysis using GEE and the innovating application of GMM to analyze pseudo-observations, comparisons of the GMM models to benchmark methods are presented through simulations in Section \ref{s:Simulation study}, and illustrations through real-data examples in Section \ref{s:application}. We conclude with some final remarks and future extensions of the proposed approach in Section \ref{s:discuss}.

\newpage
\section{Methods}
\label{s:Methods}
\subsection{Pseudo-observations computation to estimate hazard ratios}
As previously stated, pseudo-observations have been used in many applications in survival analysis. Suppose that $T_1, ..., T_n$ are $n$ independent and identically distributed time to event variables and $\widehat{\theta}$ is an unbiased estimator of a quantity of interest $\theta = \mathbb{E}(h(T_i))$, where $h$ is a known function. For individual $i$, the pseudo-observation is calculated as:        
\begin{equation}
    \widehat{\theta}_{i} = n\widehat{\theta} - (n-1)\widehat{\theta}^{-i}
\end{equation} 
where $\widehat{\theta}^{-i}$ is the value of the estimator when the $i$-th individual is removed from the data set. From this definition, $\widehat{\theta}_i$ is an approximately unbiased estimator of $\theta$. Pseudo-observations can be interpreted as an individual contribution to the overall estimate of the quantity of interest.  

In the case of estimating hazard ratios from a proportional hazard model, the pseudo-observation of the $i$-th individual at time $t_k$ is defined as:
\begin{equation}
    y_{ik} = n\widehat{S}(t_k) - (n-1)\widehat{S}^{-i}(t_k)
\end{equation}
where $\widehat{S}(t_k)$ is the Kaplan-Meier estimator of the survival probability at time $t_k$ and $\widehat{S}^{-i}(t_k)$ is the Kaplan-Meier estimator of the survival probability at time $t_k$ after removing the $i$-th individual from the data set. 

From this definition, pseudo-observations can take values around $0$ and $1$ that vary over the follow-up time depending on the status (censored / uncensored) of each patient. For all the individuals at risk at time $t_k$, their pseudo-observation is greater than one. If one individual experiences an event, the corresponding pseudo-observations will be negative for all times after this event. If one individual is censored, the pseudo-observations will always be positive and will decrease towards 0 for all times after the last event time of the data set which has occurred before its censoring time \citep{Andersen2010}. These pseudo-observations are then analyzed as an outcome variable in a generalized linear model with a cloglog link function to interpret the estimated regression coefficients as hazard ratios from a Cox model. Below is the justification for choosing this particular link function. 

Since the Cox proportional hazard model with covariates $\mathbf{X_i}$ can be written as $S(t|\mathbf{X_i}) = S_0(t)^{\exp(\boldsymbol{\beta}\mathbf{X_i})}$. Applying the complementary log-log link function $g(x) = \log(-\log(x))$ to the previous equation results in
\begin{equation}
    g(S(t|\mathbf{X_i})) = \log(H_0(t)) + \boldsymbol{\beta}\mathbf{X_i}
\end{equation}
where $H_0(t) = -\log(S_0(t))$ is the cumulative baseline hazard.
Assuming that the censoring does not depend on covariates and the event times, \citet{Graw2009} developed theoretical justifications to prove the approximate unbiasedness of pseudo-observations given the covariates (i.e. $E(y_{i}|\mathbf{X_i}) \approx S(t|\mathbf{X_i}))$. For additional information on the theoretical proprieties of pseudo-observations, refer to the discussion in \citet{Andersen2010} and to \citet{Overgaard2017}. Consequently, pseudo-observations can be analyzed as an outcome variable in the generalized linear model: 
\begin{equation}
    g(E(y_{i}|\mathbf{X_i})) = \log(H_0(t)) + \boldsymbol{\beta} \mathbf{X_i}.
\end{equation}

We can compute the pseudo-observations not only at one time point $t$ but at different time points ($t_k, k=1,\ldots,K$) for each individual. A multivariate model for $S(t_1|\mathbf{X}), \ldots, S(t_K|\mathbf{X})$ can be analyzed similarly, where $\mathbf{y_i}$ is now a $K$-dimensional vector since several pseudo-observations are defined for each individual. This model, extended for multiple time points, corresponds to a Cox model where the $\beta$'s can be interpreted as hazard ratios.
 
\subsection{Generalized Estimating Equations (GEE)}
\citet{Andersen2003} suggest analyzing pseudo-observations as an outcome variable in a regression model using the generalized estimated equations from \citet{Liang1986}. This marginal approach is based on quasi-likelihood functions where only the moments are defined \citep{McCullagh1991}.
Suppose that $\mathbf{X_i}$ = $(\mathbf{X_{i1}}, \ldots, \mathbf{X_{iK}})^T$, $\mathbf{y_i} = (y_{i1}, \ldots, y_{iK})^T$, and $\mathbf{\boldsymbol{\mu}_i} = (\mu_{i1}, \ldots, \mu_{iK})^T$ are the covariates matrix (of dimension $K \times P$), the outcome vector, and the mean vector for the $i$-th individual, respectively. The mean model is specified as:
\begin{equation} E(\mathbf{y_{i}}|\mathbf{X_i}) = \rm{cloglog}^{-1}(\beta_0 + \beta_1\mathbf{X_{i1}} + \beta_2\mathbf{X_{i2}} + \cdots + \beta_K\mathbf{X_{iK}})\end{equation}
with $\beta_0$ the intercept, $\beta_1$ the treatment effect, and $\beta_2,\ldots,\beta_K$ the time effects of the $K-1$ dummy variables, derived from the indicator of the time of which the pseudo-observation is defined. In practice, $K = 5$ time points equally spaced on the event time scale are sufficient to capture all the information from the Kaplan-Meier curve \citep{Klein2014}. The coefficient of interest in this model is the treatment effect ($\beta_1$) since the $K-1$ dummy time variables only serve as adjustment variables. Although more covariates may be added to account for other explanatory features, as illustrated in the real-data examples (see Section \ref{s:application}), for now, only the treatment effect is considered. Therefore, we note $\boldsymbol{\beta} = (\beta_0,\ldots,\beta_{K})^T$ the vector of parameters to estimate, of dimension $P = K+1$ in this particular case. 

The vector $\boldsymbol{\beta}$ is estimated by solving the score equations: 
    \begin{equation} \mathbf{U_n(\boldsymbol{\beta)}}= \frac{1}{n}\sum^n_{i = 1} \mathbf{\mathbf{u_i}(\boldsymbol{\beta)}}=\frac{1}{n}\sum^n_{i = 1} \mathbf{D_i}^T\mathbf{R}^{-1}(\alpha)(\mathbf{y_i} - \mathbf{\boldsymbol{\mu}_i})=0,\end{equation}
where $\mathbf{D_i} = \partial \mathbf{\boldsymbol{\mu}_i}/\partial\boldsymbol{\beta}^T$ is a $K \times P$ matrix and the working correlation matrix $\mathbf{R}(\alpha)$ is assumed to be of specific forms: the two common ones are the independence form where $\mathbf{R}(\alpha)$ equals the identity matrix and the exchangeable matrix defined as $1$ on the diagonal and $\alpha$ elsewhere.

The nuisance parameter $\alpha$ is estimated alternatively with $\boldsymbol{\beta}$, switching between estimating $\boldsymbol{\beta}$ for fixed values of $\widehat{\alpha}$, and estimating $\alpha$ for fixed values of $\widehat{\boldsymbol{\beta}}$. Using a consistent estimator of $\alpha$ suggested in \citet{Liang1986}, the GEE estimator $\widehat{\boldsymbol{\beta}}$ is also consistent, even if the working correlation matrix is misspecified. When applying GEE to pseudo-observations, the working correlation matrix is usually assumed to be independent, even if pseudo-observations are correlated by definition \citep{Klein2008}.
The GEE estimator converges in distribution to a normal distribution: 
\begin{equation}\sqrt{n}(\widehat{\boldsymbol{\beta}} - \boldsymbol{\beta}) \overset{d}{\to} N(0, \boldsymbol{\Gamma}),\end{equation} with $\boldsymbol{\Gamma} = \underset{n\to +\infty}{\lim} \mathbf{\boldsymbol{\Gamma}_0}^{-1}\mathbf{\boldsymbol{\Gamma}_1}\mathbf{\boldsymbol{\Gamma}_0}^{-1}$, where $\mathbf{\boldsymbol{\Gamma}_0} = \frac{1}{n}\sum_{i=1}^n \mathbf{D_i}^T \mathbf{R}^{-1}(\alpha) \mathbf{D_i}$, and 
\newline $\mathbf{\boldsymbol{\Gamma}_1} = \frac{1}{n}\sum_{i=1}^n \mathbf{D_i}^T \mathbf{R}^{-1}(\alpha) (\mathbf{y_i} - \mathbf{\boldsymbol{\mu_i}})(\mathbf{y_i} - \mathbf{\boldsymbol{\mu_i}})^T \mathbf{R}^{-1}(\alpha)\mathbf{D_i}$. The estimator of the $\boldsymbol{\Gamma}$ matrix, referred to as the sandwich or robust variance estimator, is obtained by evaluating the matrices $\mathbf{\boldsymbol{\Gamma}_0}$ and $\mathbf{\boldsymbol{\Gamma}_1}$ at their empirical estimates. \citet{Jacobsen2016} have shown that this estimator is slightly conservative, but the correction of the variance proposed by the authors is numerically small. Therefore, the sandwich estimator is still commonly used \citep{Bouaziz2023}.

\subsection{Frequentist Generalized Method of Moments}
 The generalized method of moments (GMM) is defined by \citet{Hansen1982} and is widely used in econometrics contrary to biostatistics. \citet{Ziegler1995} showed that the GMM and the GEE approaches give asymptotically equivalent estimators. The principle of GMM is to combine multiple moments through score equations. The system of equations becomes over-identified as the number of equations exceeds the number of unknown parameters. Therefore, the exact solution cannot be found anymore. The estimates are then found by minimizing an objective function defined using the score vector and a weight matrix that gives more weights to the equations with less variability. 
 
 \citet{Qu2000} proposed a GMM approach for longitudinal data with a theoretical efficiency improvement under correlation misspecification. In this particular case, only the first moment (ie. the mean model $\mathbf{\boldsymbol{\mu_i}}$) is specified identically to GEE. This approach can be viewed as an extension of GEE since the general idea is to express the inverse of the working correlation matrix, $\mathbf{R}$, as a linear combination of $J$ basis matrices, $\mathbf{R}^{-1} \approx \sum^J_{j = 1} a_j\mathbf{M_j}$. The inverses of the different working correlation matrices specified in the GEE approach can be expressed as a sum of the basis matrices. For example, the inverse of the independence matrix is expressed as $\mathbf{R}^{-1} = a_1\mathbf{M_1}$, the inverse of the exchangeable matrix as $\mathbf{R}^{-1} = a_1\mathbf{M_1} + a_2\mathbf{M_2}$, where $\mathbf{M_1}$ is the identity matrix and $\mathbf{M_2}$ is the matrix with $0$ on the diagonal and $1$ elsewhere. The first-order auto-regressive (AR-1) working correlation matrix is defined with coefficients $r_{ij} = \alpha^{|i-j|}$, with $i$ the line and $j$ the column number. Its inverse $\mathbf{R}^{-1}$ can be approximated by two working correlation matrices: $\mathbf{M_1}$ is the identity, and $\mathbf{M_2}$ is the matrix with $1$ on the two diagonals on both sides of the main diagonal and $0$ elsewhere.
 
With the GMM approach, $\mathbf{u_i(\boldsymbol{\beta})}$ is now a $(J\times P)$-dimensional score vector defined as \begin{equation}\mathbf{u_i(\boldsymbol{\beta})} = \begin{Bmatrix}
\mathbf{D_i}^T\mathbf{M_1}(\mathbf{y_i} - \mathbf{\boldsymbol{\mu_i}})\\
\mathbf{D_i}^T\mathbf{M_2}(\mathbf{y_i} - \mathbf{\boldsymbol{\mu_i}})\\
\vdots\\
\mathbf{D_i}^T\mathbf{M_J}(\mathbf{y_i} - \mathbf{\boldsymbol{\mu_i}})
\end{Bmatrix},\end{equation}

and the objective function (quadratic inference function) is written as 
\begin{equation}Q_n(\boldsymbol{\beta}) = \mathbf{U_n}^T\mathbf{(\boldsymbol{\beta})C_n}^{-1}\mathbf{(\boldsymbol{\beta})U_n(\boldsymbol{\beta})},\end{equation}
where $\mathbf{U_n(\boldsymbol{\beta})} = \frac{1}{n}\sum_{i = 1}^n \mathbf{u_i(\boldsymbol{\beta})}$, and $\mathbf{C_n(\boldsymbol{\beta})} = \frac{1}{n^2}\sum_{i = 1}^n \mathbf{u_i(\boldsymbol{\beta})}\mathbf{u_i}^T(\boldsymbol{\beta})$. Contrary to GEE, the vector $\mathbf{U_n(\boldsymbol{\beta})}$ now contains more equations than unknown parameters, and the $\boldsymbol{\beta}$'s are estimated by minimizing the quadratic inference function $\widehat{\boldsymbol{\beta}} = \rm{argmin}(Q_n(\boldsymbol{\beta}))$. The Newton-Raphson algorithm can be used to minimize this function, with starting values usually chosen to be the least squared estimates. 

A consistent variance estimator can also be derived with a sandwich form:
\begin{equation}\mathbf{\widehat{\rm{cov}}(\widehat{\boldsymbol{\beta}})} = \frac{1}{n} \left[\{\partial \mathbf{U_n(\widehat{\boldsymbol{\beta}})}^T/\partial\boldsymbol{\beta}\}\mathbf{C_n}^{-1}\{\partial \mathbf{U_n(\widehat{\boldsymbol{\beta}})}/\partial\boldsymbol{\beta}^T\}\right]^{-1} .\end{equation}

Under some regulatory conditions, \citet*{Yu2020} have shown that this approach produces identical point estimates compared to GEE and robust covariances with an independence or exchangeable working matrix. Regarding the analysis of pseudo-observations with GMM, the mean model is identical to the one of the GEE approach, with a cloglog link function to interpret the regression coefficients as hazard ratios.

\subsection{Bayesian Generalized Method of Moments}
\subsubsection{Model}
The formulation of the Bayesian generalized method of moments can be derived by considering that the minimization problem of the GMM can be converted to a Bayesian sampling problem. By applying the Central Limit Theorem, 
\begin{equation}\mathbf{U_n(\boldsymbol{\beta})} \overset{d}{\longrightarrow} N(0,\boldsymbol{\Sigma}(\boldsymbol{\beta})), \text{ as $n \to \infty$}\end{equation}

where $\boldsymbol{\Sigma}(\boldsymbol{\beta}) = \underset{n \to \infty}{\lim}\mathbf{\boldsymbol{\Sigma}_n}(\boldsymbol{\beta}),$ then
\begin{equation}Q_n(\boldsymbol{\beta}) \overset{d}{\longrightarrow} \chi_{(J-1) \times P}^2.\end{equation}
A chi-squared test can be derived, analogous to the usual likelihood ratio test, where $Q_n(\boldsymbol{\beta})$ behaves like $-2\log L(y|\boldsymbol{\beta})$ with $L(y|\boldsymbol{\beta})$ being the likelihood function \citep{Hansen1982}. Thus, the GMM approximates the likelihood for selected moments of the data without specifying the full likelihood \citep{Chernozhukov2003}. 

Given these theoretical results, \citet{Yin2009} presented a Bayesian version of GMM by defining a pseudo-likelihood function as follows: \begin{equation}\tilde{L}(y| \boldsymbol{\beta}) \propto \exp\{-\frac{1}{2}\mathbf{U_n}^T(\boldsymbol{\beta})\mathbf{\boldsymbol{\Sigma}_n}^{-1}(\boldsymbol{\beta})\mathbf{U_n(\boldsymbol{\beta})}\},\end{equation}
with $\mathbf{\boldsymbol{\Sigma}_n}(\boldsymbol{\beta}) =  \frac{1}{n^2}\sum_{i = 1}^n \mathbf{u_i(\boldsymbol{\beta})}\mathbf{u_i}^T(\boldsymbol{\beta}) - \frac{1}{n}\mathbf{U_n(\boldsymbol{\beta})}\mathbf{U_n}^T(\boldsymbol{\beta})$. Note that $\mathbf{\boldsymbol{\Sigma}_n}^{-1}(\boldsymbol{\beta})$ in the quasi-likelihood has an additional term compared to the empirical covariance matrix $\mathbf{C_n}(\boldsymbol{\beta})$ in the quadratic inference function $Q_n(\boldsymbol{\beta})$.

\citet{Yin2009} showed the validity of the posterior distribution resulting from this pseudo-likelihood. However, this pseudo-likelihood function, $\tilde{L}(y| \boldsymbol{\beta})$, is only defined on the support of $\mathbb{R}^P$ where $\mathbf{\boldsymbol{\Sigma}_n}$ is invertible, which is restricted due to the cloglog link function used for pseudo-observations analysis. For example, Figure~\ref{orsini:fig1} represents the pseudo-likelihood function as a function of the treatment effect $\beta_1$; all other parameters are fixed at their GEE estimates. The gray zone indicates the values of $\beta_1$ for which the matrix $\mathbf{\boldsymbol{\Sigma}_n}$ is not invertible. 
Thus, convergence issues may occur when parameter values fall outside this local support. The inverse link function being $x \to  \exp(-\exp(x))$ may result in extreme values of the parameters. In practice, when the Bayesian sampler draws a value of $\boldsymbol{\beta}$ far from the true value, $\mathbf{\boldsymbol{\Sigma}_n}$ becomes non-invertible. 
Consequently, it is essential to calibrate the Bayesian algorithm well by choosing appropriate priors and starting values for each model parameter. Below, we specify (i) how to choose appropriate prior distributions and (ii) the algorithm to generate sensible starting values.

\subsubsection{Choosing appropriate priors}
\label{s:priors}
\begin{figure*}[!ht]\centering
\includegraphics[width=15cm]{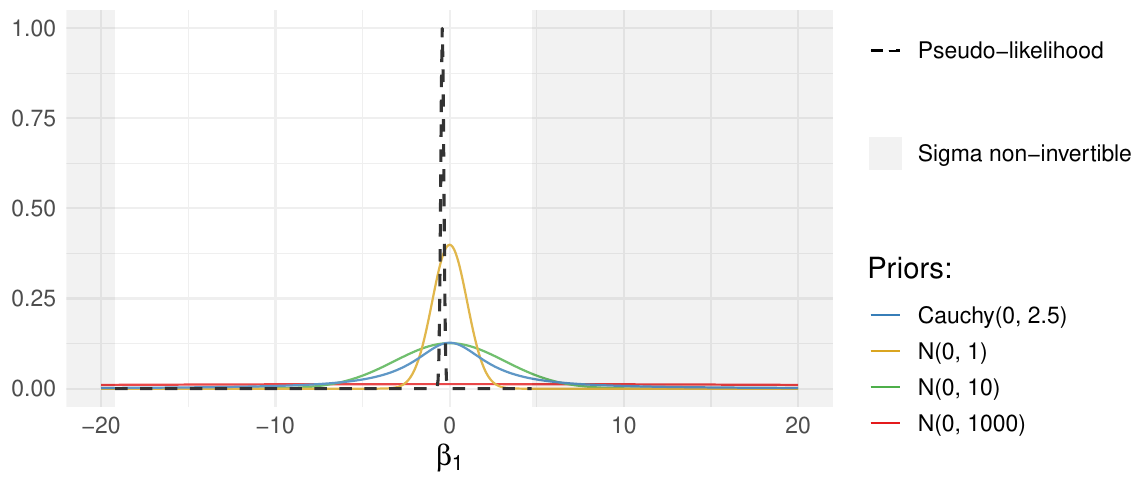}
\caption{Example of the pseudo-likelihood function (black dashed line) depending on the treatment effect ($\beta_1$), all the other parameters are fixed to their GEE estimations. Solid lines represent different priors that have been investigated. Gray zone represents values $\beta_1$ where $\mathbf{\boldsymbol{\Sigma}_n}$ is non-invertible and thus, the likelihood function is not defined.}
\label{orsini:fig1}
\end{figure*}

Choosing appropriate priors for the cloglog scale partially resolves the convergence issue previously mentioned. \citet{Gelman2008} proposed to use Cauchy(0, 2.5) for all regression coefficients as default priors in generalized linear regression models after centering and re-scaling all the input variables. These weakly informative priors reflect the fact that large changes on the logit or cloglog scale are rare. Using weak Gaussian priors such as $N(0,10)$ or $N(0,1)$, as recommended by the \citet{StanDev2020}, can provide an alternative to Cauchy priors. They may be more adapted to the pseudo-likelihood defined on a small support because they have lighter distribution tails (See Figure~\ref{orsini:fig1}). We do not recommend using extremely vague priors, for example, $N(0,1000)$, as they correspond to unrealistic values on the probability scale. As we estimate hazard ratios, such large priors are unreasonable. Although weak priors are more informative than flat priors, they are vague enough compared to the pseudo-likelihood \citep*{Gelman2017}.

\subsubsection{Starting values}
\label{s:startingvalues}
Setting starting values randomly is not optimal as they might fall outside the definition support of the pseudo-likelihood. Even if they are on the edge of the support, the $\boldsymbol{\beta}$ values may fall outside the support after a few iterations, especially during the warm-up period where the No-U-turn sampler (NUTS) chooses the step size adaptively \citep{Hoffman2014}. Consequently, the step size may be too large for some iterations, resulting in poor convergence. To overcome this issue, we propose to generate starting values of the NUTS in a similar manner to the one used for the frequentist GMM, which are based on the least square estimates, while taking into account the cloglog link function. The generation of these initial values includes three steps described as follows:

\begin{enumerate}
    \item Truncate the pseudo-observations to $[\epsilon, 1-\epsilon]$ where $\epsilon>0$ takes a small value. Pseudo-observations above $1-\epsilon$ are set to $1-\epsilon$, and pseudo-observations below $\epsilon$ are set to $\epsilon$. This step is needed to apply the cloglog function to pseudo-observations as it is defined on $]0,1[$.
    \item Apply the cloglog function to the truncated pseudo-observations.
    \item Perform a linear regression model using these modified pseudo-observations as a continuous outcome and treatment factor and dummy time variables as covariates, then use the ordinary least square estimates as starting values.
\end{enumerate} 

Different values of the truncation parameter $\epsilon$ can be chosen, resulting in different starting values for each chain of the NUTS sampling. We emphasize the point that this process does not give correct estimation (1) the pseudo-observations have been truncated to $[\epsilon, 1-\epsilon]$ which may induce a bias in the estimates, (2) the cloglog is applied to the pseudo-observations themselves and not to the mean vector, and (3) the correlation between pseudo-observations of the same individual is not taken into account using the least square estimation. So, this is only a generic and straightforward process to generate starting values in the definition support of the pseudo-likelihood to improve the convergence.

\section{Simulation study}
\label{s:Simulation study}
We performed a simulation study to assess the performance of the frequentist and Bayesian GMM applied to pseudo-observations in order to estimate the hazard ratio of the treatment effect. The purpose of this simulation study was (a) to assess the validity of the pseudo-observations analysis using the frequentist and Bayesian GMM; (b) to compare the performances of the GMM models to the three benchmark methods: the Cox proportional hazard model, and the GEE approach based on pseudo-observations in the frequentist framework, and to the piecewise exponential model in the Bayesian framework; and (c) to evaluate the impact on the estimation of the different choices made in the pseudo-observations based models, i.e., the number of time points and the form of the working correlation matrix.

\subsection{Settings}
Simulations were based on a two-arm randomized clinical trial with a time-to-event outcome. Event times were generated from a Weibull distribution $f(t|a,b) = (a/b)(t/b)^{a-1}\exp((t/b)^a)$ with shape parameter $a=0.6$ and scale parameter $b = \exp(- \frac{\beta_1X_1}{a})$ depending on the treatment indicator $X_1$ coded 1 for experimental arm and 0 for control arm. This corresponds to a randomized control trial with a median survival time of approximately 6 months in the control arm for the core scenario (i.e. with $n=500$, a censoring rate of $20\%$, and $\log \rm{HR}=-0.3$). No other explanatory variables were considered in the simulations. Censoring times were generated independently following a uniform distribution. The parameter of the uniform distribution was chosen according to the desired censoring rate, following \citet{Wan2017}. The simulation parameters were the sample size, varying from 50 to 1000, the censoring rate, ranging from 5\% to 70\%, and the true treatment effect varying from -0.5 to -0.1 (log scale) corresponding to hazard ratio from 0.6 to 0.9, approximately. These specifications represent different scenarios of randomized clinical trials with different sizes of the treatment effect between the experimental and control arms. Bias, average standard error (ASE), root-mean-square error (RMSE), and coverage rate from $95\%$ equal-tailed intervals were calculated from $n_{\rm{sim}} = 1000$ replications for each scenario. 

All computations were performed using the R Language for Statistical Computing  (\citet{RCoreTeam2021}, version 4.1.2). Pseudo-observations have been computed using the R package \texttt{pseudo} \citep*{PoharPerme2017}, with $K = 5$ time points. Because the R package \texttt{qif} by \citet*{Jiang2019}, only allows the use of the canonical links (identity, log, logit, inverse), we developed an R script to implement the frequentist GMM with a cloglog link function.

We also developed a specific script to implement the Bayesian GMM using the Stan software \citep{Carpenter2017}. The model was then compiled via the \texttt{rstan} R package, \citep{StanDev2023}. The NUTS sampling was performed with $3$ chains of each $5000$ iterations after a warm-up of $1000$ iterations, and thinning of $5$, yielding $3000$ iterations overall. As mentioned in Section \ref{s:Methods}, weakly informative priors were specified for all parameters (intercept, treatment effect, and dummy time variables). In scenarios with $n = 50$ or $n = 100$, a prior distribution of $N(0, 1)$ was specified for all parameters; in all the other scenarios, $N(0,10)$ prior was specified. Initial parameters were set by fixing the truncation parameters $\epsilon \in (0.01, 0.05, 0.1)$ for the three chains, respectively. The convergence diagnoses were performed through trace plots checking and $\widehat{R}$ estimation \citep{Vehtari2021}. 

The R package \texttt{geepack} was used to implement the GEE approach on pseudo-observations \citep{Højsgaard2022}. Multiple jackknife variance estimators are given in addition to the sandwich variance estimator. The approximate jackknife variance estimator is recommended to analyze pseudo-observations, following suggestions in \citet{Klein2008}. All the estimators are equivalent for large samples, as referenced in \citet{Yan2004}. We used the \texttt{spBayesSurv} R package by \citet*{Zhou2020} to implement the Bayesian piecewise exponential model. The number of intervals for the time partition was chosen according to the number of events following the rule in \citet{Murray2014} (i.e., $M = \max\{5, \min(\frac{r}{8}, 20)\}$) where $M$ is the total number of intervals, and $r$ is the observed number of events in the trial data set. The baseline hazard is assumed constant within each interval: $h_0(t) = \sum_{m=1}^M h_m I\{t \in I_m\}$. All the priors were kept as default, i.e., the priors for the baseline hazard were $h_m \sim \Gamma(1, \widehat{h})$ with $\widehat{h}$ the maximum likelihood estimate of the rate parameter from fitting an exponential proportional hazard model, and the priors for the log hazard ratio was $\beta_1 \sim N(0,10^5)$.

\subsection{Results}
Table~\ref{orsini:tab1} represents the estimates of the hazard ratio (on a log scale) for different scenarios with a substantial treatment effect of $HR=0.74$ ($\log HR=-0.3$), a censoring rate of $20\%$ and different sample sizes. Overall, GMM approaches (frequentist and Bayesian) produce valid inferences with a bias that decreases toward zero as the sample size increases. From small and moderate sample sizes ($n=50$, $100$, and $200$), Bayesian GMM results in slightly higher bias (varying from $-0.0852$ for $n=50$ to $-0.0155$ for $n=200$) compared to frequentist GMM (varying from $-0.0149$ to $0.0004$) and similar standard errors. The coverage rates are close to the nominal coverage rate of $95\%$ for large sample sizes ($n\geq 500$). When comparing these performances with the ones of the three benchmark models: the Cox model, the pseudo-observations-based GEE model, and the piecewise exponential model, GMM gives similar results (bias close to zero, similar standard errors and RMSEs, coverage rate close to $95\%)$. 
We note, however, a slight difference for the scenarios with small sample sizes ($n\leq 100$). For these scenarios, as expected, frequentist GMM and GEE give similar results with a higher variance than the estimates of the Cox and Bayesian exponential piecewise models. For example, the average standard error is $0.257$ for frequentist GMM compared to $0.228$ for the Cox model for $n=100$. This result is consistent with \citet{Andersen2003}. This results in a higher RMSE for pseudo-observations-based models. The estimates from Bayesian methods are more biased than the frequentist approaches, especially for $n=50$ with a higher bias for the Bayesian GMM ($-0.0852$ for Bayesian GMM, $-0.0574$ for piecewise exponential model versus $-0.0203$ for Cox). In addition, this bias decreases when the treatment effect decreases (toward 0) for the piecewise exponential model, while it remains constant for the Bayesian GMM when $n\leq 100$.

\begin{table}[!ht]\centering
\caption{Performances of the frequentist and Bayesian GMM compared to the Cox, GEE, and piecewise exponential model (PEM) with a  true log hazard ratio of -0.3 (HR=0.74), a censoring rate of $20\%$ and different sample sizes.}
\label{orsini:tab1}
\medskip
\begin{tabular}{llrrrr}
\toprule
$n$ &Methods&Bias&ASE$^1$&RMSE$^2$&Coverage\\ %
\midrule

\addlinespace[0.3em]
\textbf{50}&\multicolumn{5}{l}{\textbf{Frequentist}}\\
&\hspace{1em}Cox & -0.0203 & 0.326 & 0.332 & 95.2\\
&\hspace{1em}GEE & -0.0149 & 0.355 & 0.385 & 92.6\\
&\hspace{1em}GMM & -0.0149 & 0.367 & 0.385 & 93.5\\
\addlinespace[0.3em]
&\multicolumn{5}{l}{\textbf{Bayesian}}\\
&\hspace{1em}PEM & -0.0574 & 0.332 & 0.367 & 92.7\\
&\hspace{1em}GMM & -0.0852 & 0.354 & 0.386 & 91.9\\
\bottomrule

\addlinespace[0.3em]
\textbf{100}& \multicolumn{5}{l}{\textbf{Frequentist}}\\
&\hspace{1em}Cox & 0.0054 & 0.228 & 0.245 & 93.2\\
&\hspace{1em}GEE & 0.0100 & 0.253 & 0.270 & 93.6\\
&\hspace{1em}GMM & 0.0100 & 0.257 & 0.270 & 94.0\\
\addlinespace[0.3em]
&\multicolumn{5}{l}{\textbf{Bayesian}}\\
&\hspace{1em}PEM & -0.0178 & 0.233 & 0.264 & 91.0\\
&\hspace{1em}GMM & -0.0341 & 0.253 & 0.273 & 92.8\\
\bottomrule

\addlinespace[0.3em]
\textbf{200}&\multicolumn{5}{l}{\textbf{Frequentist}}\\
&\hspace{1em}Cox & 0.0003 & 0.160 & 0.161 & 94.1\\
&\hspace{1em}GEE & 0.0004 & 0.180 & 0.188 & 93.5\\
&\hspace{1em}GMM & 0.0004 & 0.181 & 0.188 & 93.5\\
\addlinespace[0.3em]
&\multicolumn{5}{l}{\textbf{Bayesian}}\\
&\hspace{1em}PEM & -0.0202 & 0.163 & 0.174 & 93.4\\
&\hspace{1em}GMM & -0.0155 & 0.189 & 0.195 & 93.1\\
\bottomrule

\addlinespace[0.3em]
\textbf{500}& \multicolumn{5}{l}{\textbf{Frequentist}}\\
&\hspace{1em}Cox & 0.0028 & 0.101 & 0.100 & 95.0\\
&\hspace{1em}GEE & 0.0032 & 0.114 & 0.112 & 95.4\\
&\hspace{1em}GMM & 0.0032 & 0.114 & 0.112 & 95.5\\
\addlinespace[0.3em]
&\multicolumn{5}{l}{\textbf{Bayesian}}\\
&\hspace{1em}PEM & -0.0063 & 0.102 & 0.104 & 94.0\\
&\hspace{1em}GMM & -0.0028 & 0.116 & 0.113 & 95.4\\
\bottomrule

\addlinespace[0.3em]
\textbf{1000}&\multicolumn{5}{l}{\textbf{Frequentist}}\\
&\hspace{1em}Cox & 0.0032 & 0.071 & 0.072 & 94.8\\
&\hspace{1em}GEE & 0.0005 & 0.081 & 0.082 & 95.2\\
&\hspace{1em}GMM & 0.0005 & 0.081 & 0.082 & 95.2\\
\addlinespace[0.3em]
&\multicolumn{5}{l}{\textbf{Bayesian}}\\
&\hspace{1em}PEM & -0.0017 & 0.071 & 0.073 & 94.9\\
&\hspace{1em}GMM & -0.0026 & 0.082 & 0.082 & 95.2\\
\bottomrule
\multicolumn{6}{l}{\rule{0pt}{1em}\textsuperscript{1} ASE = Average Standard Error}\\
\multicolumn{6}{l}{\rule{0pt}{1em}\textsuperscript{2} RMSE = Root Mean Square Error}\\
\addlinespace[1em]
\end{tabular}
\end{table}

When varying the censoring rate for the core scenario ($\log \rm{HR}=-0.3$, $\rm{HR}=0.74$ and $n=500$), the performances of GMM (frequentist and Bayesian) are similar to the three benchmark methods (Table~\ref{orsini:tab2}). The more pronounced differences between these pseudo-observation-based approaches (GEE and GMM) and the Cox and Bayesian piecewise exponential methods occur with higher average standard error and RMSE for large censoring rates ($30\%$ and $70\%$). For example, the average standard error of the Bayesian GMM is $0.194$ compared to $0.166$ for the Bayesian piecewise exponential model for a censoring rate of $70\%$.

\begin{table}[!ht]\centering
\caption{Performances of the frequentist and Bayesian GMM compared to the Cox, GEE, and piecewise exponential model (PEM) with a true log hazard ratio of -0.3 (HR=0.74), different censoring rates, and a sample size of 500.}
\label{orsini:tab2}
\medskip
\begin{tabular}{llrrrr}
\toprule
CR$^1$&Methods&Bias&ASE$^2$&RMSE$^3$&Coverage\\
\midrule

\addlinespace[0.3em]
\textbf{5\%}&\multicolumn{5}{l}{\textbf{Frequentist}}\\
&\hspace{1em}Cox & 0.0016 & 0.093 & 0.094 & 94.0\\
&\hspace{1em}GEE & 0.0032 & 0.107 & 0.105 & 95.1\\
&\hspace{1em}GMM & 0.0032 & 0.107 & 0.105 & 95.1\\
\addlinespace[0.3em]
&\multicolumn{5}{l}{\textbf{Bayesian}}\\
&\hspace{1em}PEM & -0.0080 & 0.094 & 0.097 & 93.6\\
&\hspace{1em}GMM & -0.0036 & 0.109 & 0.108 & 95.1\\
\bottomrule

\addlinespace[0.3em]
\textbf{10\%}&\multicolumn{5}{l}{\textbf{Frequentist}}\\
&\hspace{1em}Cox & 0.0013 & 0.095 & 0.096 & 94.4\\
&\hspace{1em}GEE & 0.0032 & 0.109 & 0.107 & 95.5\\
&\hspace{1em}GMM & 0.0032 & 0.109 & 0.107 & 95.5\\
\addlinespace[0.3em]
&\multicolumn{5}{l}{\textbf{Bayesian}}\\
&\hspace{1em}PEM & -0.0080 & 0.096 & 0.098 & 94.0\\
&\hspace{1em}GMM & -0.0032 & 0.111 & 0.108 & 95.3\\
\bottomrule

\addlinespace[0.3em]
\textbf{20\%}&\multicolumn{5}{l}{\textbf{Frequentist}}\\
&\hspace{1em}Cox & 0.0028 & 0.101 & 0.100 & 95.0\\
&\hspace{1em}GEE & 0.0032 & 0.114 & 0.112 & 95.4\\
&\hspace{1em}GMM & 0.0032 & 0.114 & 0.112 & 95.5\\
\addlinespace[0.3em]
&\multicolumn{5}{l}{\textbf{Bayesian}}\\
&\hspace{1em}PEM & -0.0063 & 0.102 & 0.104 & 94.0\\
&\hspace{1em}GMM & -0.0028 & 0.116 & 0.113 & 95.4\\
\bottomrule

\addlinespace[0.3em]
\textbf{30\%}& \multicolumn{5}{l}{\textbf{Frequentist}}\\
&\hspace{1em}Cox & 0.0039 & 0.108 & 0.107 & 94.8\\
&\hspace{1em}GEE & 0.0021 & 0.121 & 0.119 & 95.0\\
&\hspace{1em}GMM & 0.0021 & 0.121 & 0.119 & 95.2\\
\addlinespace[0.3em]
&\multicolumn{5}{l}{\textbf{Bayesian}}\\
&\hspace{1em}PEM & -0.0055 & 0.109 & 0.111 & 94.0\\
&\hspace{1em}GMM & -0.0041 & 0.123 & 0.120 & 95.1\\
\bottomrule

\addlinespace[0.3em]
\textbf{70\%}& \multicolumn{5}{l}{\textbf{Frequentist}}\\
&\hspace{1em}Cox & 0.0018 & 0.165 & 0.165 & 94.3\\
&\hspace{1em}GEE & 0.0006 & 0.184 & 0.185 & 94.9\\
&\hspace{1em}GMM & 0.0006 & 0.185 & 0.185 & 94.9\\
\addlinespace[0.3em]
&\multicolumn{5}{l}{\textbf{Bayesian}}\\
&\hspace{1em}PEM & -0.0068 & 0.166 & 0.170 & 93.9\\
&\hspace{1em}GMM & -0.0119 & 0.194 & 0.188 & 95.0\\
\bottomrule
\multicolumn{6}{l}{\rule{0pt}{1em}\textsuperscript{1} CR = Censoring Rate}\\
\multicolumn{6}{l}{\rule{0pt}{1em}\textsuperscript{2} ASE = Average Standard Error}\\
\multicolumn{6}{l}{\rule{0pt}{1em}\textsuperscript{3} RMSE = Root Mean Square Error}\\
\addlinespace[1em]
\end{tabular}
\end{table}

We evaluated the impact of different effect sizes from small to important (HR=0.90, 0.74, and 0.60) on the performances of the frequentist and Bayesian GMM approaches for the core scenario ($n=500$, censoring rate $=20\%$). Performances are similar to the three benchmark methods (Table~\ref{orsini:tab3}). The supporting information Tables S1a, S1b, S2a, and S2b show the performances in all the other scenarios. For a given censoring rate and sample size, the size of the treatment effect did not affect the performances of all the models.

\begin{table}[!ht]\centering
\caption{Performances of the frequentist and Bayesian GMM compared to the Cox, GEE, and piecewise exponential model (PEM) with different treatment effects (log hazard ratio of -0.1 (HR=0.9), -0.3 (HR=0.74), and -0.5 (HR=0.60), a censoring rate of $20\%$ and a sample size of $500$.}
\label{orsini:tab3}
\medskip
\begin{tabular}{llrrrr}
\toprule
HR$^1$&Methods&Bias&ASE$^2$&RMSE$^3$&Coverage\\
\midrule

\addlinespace[0.3em]
\textbf{0.9}&\multicolumn{5}{l}{\textbf{Frequentist}}\\
&\hspace{1em}Cox & 0.0029 & 0.100 & 0.100 & 94.1\\
&\hspace{1em}GEE & 0.0041 & 0.114 & 0.112 & 94.9\\
&\hspace{1em}GMM & 0.0041 & 0.114 & 0.112 & 95.0\\
\addlinespace[0.3em]
&\multicolumn{5}{l}{\textbf{Bayesian}}\\
&\hspace{1em}PEM & -0.0006 & 0.101 & 0.103 & 94.0\\
&\hspace{1em}GMM & 0.0012 & 0.116 & 0.113 & 94.9\\
\bottomrule

\addlinespace[0.3em]
\textbf{0.74}& \multicolumn{5}{l}{\textbf{Frequentist}}\\
&\hspace{1em}Cox & 0.0028 & 0.101 & 0.100 & 95.0\\
&\hspace{1em}GEE & 0.0032 & 0.114 & 0.112 & 95.4\\
&\hspace{1em}GMM & 0.0032 & 0.114 & 0.112 & 95.5\\
\addlinespace[0.3em]
&\multicolumn{5}{l}{\textbf{Bayesian}}\\
&\hspace{1em}PEM & -0.0063 & 0.102 & 0.104 & 94.0\\
&\hspace{1em}GMM & -0.0028 & 0.116 & 0.113 & 95.4\\
\bottomrule

\addlinespace[0.3em]
\textbf{0.6}&\multicolumn{5}{l}{\textbf{Frequentist}}\\
&\hspace{1em}Cox & 0.0034 & 0.102 & 0.101 & 94.8\\
&\hspace{1em}GEE & 0.0020 & 0.115 & 0.112 & 95.0\\
&\hspace{1em}GMM & 0.0020 & 0.115 & 0.112 & 95.1\\
\addlinespace[0.3em]
&\multicolumn{5}{l}{\textbf{Bayesian}}\\
&\hspace{1em}PEM & -0.0107 & 0.102 & 0.104 & 94.1\\
&\hspace{1em}GMM & -0.0071 & 0.118 & 0.114 & 95.6\\
\bottomrule

\multicolumn{6}{l}{\rule{0pt}{1em}\textsuperscript{1} HR = Hazard ratio}\\
\multicolumn{6}{l}{\rule{0pt}{1em}\textsuperscript{2} ASE = Average Standard Error}\\
\multicolumn{6}{l}{\rule{0pt}{1em}\textsuperscript{3} RMSE = Root Mean Square Error}\\
\addlinespace[0.5em]
\end{tabular}
\end{table}

No convergence issue was observed through all scenarios and replicates, with $\widehat R$ close to 1. The Bayesian GMM was run with a parallelized code using a server HPE DL385 (2.0 GHz) with $150$ virtual cores. In the core scenario (with $n = 500$ patients), the pseudo-observations computation took less than 1 second and the median running time of one chain was $12$ minutes.

\subsection{Sensitivity analysis on the number of time points}

When computing pseudo-observations, the choice of the time points, $K$, remains arbitrary. Some authors (\citet{Klein2005}, and \citet{Andersen2003}) suggested that $K=5$ is sufficient to obtain asymptotically unbiased estimates and, therefore, this value has been considered as the default. Intuitively, one wants to choose time points equally spaced on the event-times scale to capture most of the information from the Kaplan-Meier estimate. To assess the impact of the number of time points, we transformed survival data generated from the core scenario (a true log hazard ratio of $-0.3$ ($\rm{HR}=0.74$), a censoring rate of $20\%$ and a sample size of $500$) into $K=5$, $7$, and $10$ pseudo-observations separately. We analyzed the transformed data with the GEE, the frequentist GMM, and the Bayesian GMM models.

\begin{figure}[!ht]\centering
\includegraphics[width=16.5cm]{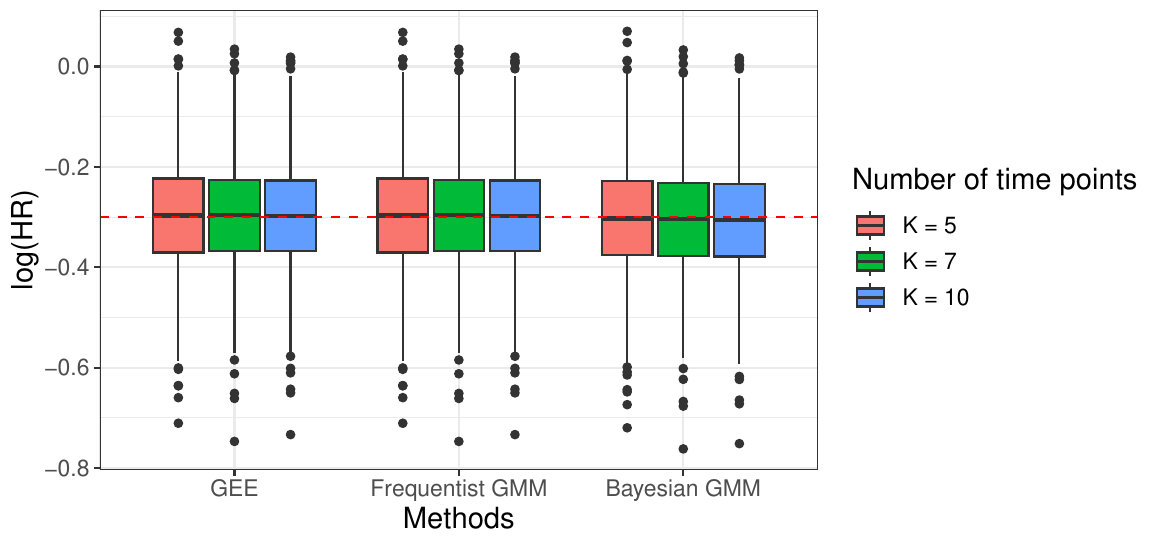}
\caption{Sensitivity analysis on the number of time points ($K=5$, $7$, and $10$) used for the computation of pseudo-observations. Box plots represent the hazard ratio (log scale) estimated from GEE, frequentist, and Bayesian GMM from 1000 replicates, with a true log hazard ratio of $-0.3$ ($HR=0.74$), a censoring rate of $20\%$ and a sample size of $500$. The horizontal red line represents the true log hazard ratio.}
\label{orsini:fig2}
\end{figure}

Figure~\ref{orsini:fig2} shows that increasing the number of time points had no impact on the median of the log hazard ratios estimated from 1000 replicates and a minor impact on the variability of these estimates, whatever the method. Supporting information Table S3 details the different performances of each model with different numbers of time points. Hence, our findings align with the previous sensitivity analysis from the literature. Thus, using $K>5$ time points is not recommended as the gain in efficiency is negligible compared to the complexity induced and the increase of the running time of the NUTS algorithm for the Bayesian GMM (data not shown).

\subsection{Sensitivity analysis on the choice of the working correlation matrix}
The choice of the working correlation matrix is one of the assumptions required when applying GEE or GMM. Previous results from pseudo-observations-based approaches are obtained with the independent working correlation matrix. This structure is often chosen in practice when using GEE to analyze pseudo-observations because the GEE method yields unbiased estimates even when the working correlation matrix is misspecified. The frequentist GMM has been developed to produce more efficient estimates than GEE for longitudinal continuous data when the working correlation matrix is misspecified \citep{Qu2000}. Thus, we analyzed the impact of different correlation matrices on hazard ratio estimates for the GEE, frequentist, and Bayesian GMM based on pseudo-observations. We firstly limit this analysis to the core scenario (true log hazard ratio of $-0.3$, censoring rate of $20\%$ and $n=500$). Table~\ref{orsini:tab4} shows that GMM approaches produce unbiased estimates whatever the working matrix and similar standard errors between the three structures. Similar results are obtained with different treatment effects (See supporting information Tables S4a and S4b). In this context, the differences in the precision of the estimations between the GEE and GMM approaches were marginal. These results concord with \citet{Yu2020}, who compared the GEE and the frequentist GMM approaches to analyze longitudinal outcomes in randomized clinical trials.

\begin{table}[!ht]\centering
\caption{Comparison of the performances of GEE and GMM models with different working correlation matrices: Independence (IND), Exchangeable (EXCH) and first-order auto-regressive (AR-1) for a true log hazard ratio of -0.3 (HR=0.74), a censoring rate of $20\%$ and sample size of $500$.}
\label{orsini:tab4}
\medskip
\begin{tabular}{llrrrr}
\toprule
Methods & WCM$^1$ & Bias & ASE$^2$ & RMSE$^3$ & Coverage\\
\midrule
\addlinespace[0.3em]
\multicolumn{6}{l}{\textbf{Frequentist}}\\
\hspace{1em}GEE & IND & 0.0032 & 0.114 & 0.112 & 95.4\\
\hspace{1em}GEE & EXCH & 0.0021 & 0.113 & 0.112 & 95.1\\
\hspace{1em}GEE & AR-1 & 0.0024 & 0.111 & 0.110 & 95.5\\
\hspace{1em}GMM & IND & 0.0032 & 0.114 & 0.112 & 95.5\\
\hspace{1em}GMM & EXCH & 0.0001 & 0.111 & 0.111 & 95.3\\
\hspace{1em}GMM & AR-1 & -0.0017 & 0.111 & 0.112 & 95.3\\
\addlinespace[0.3em]
\multicolumn{6}{l}{\textbf{Bayesian}}\\
\hspace{1em}GMM & IND & -0.0028 & 0.116 & 0.113 & 95.4\\
\hspace{1em}GMM & EXCH & -0.0055 & 0.113 & 0.113 & 95.0\\
\hspace{1em}GMM & AR-1 & -0.0073 & 0.113 & 0.113 & 94.7\\
\bottomrule
\multicolumn{6}{l}{\rule{0pt}{1em}\textsuperscript{1} WCM = Working Correlation Matrix}\\
\multicolumn{6}{l}{\rule{0pt}{1em}\textsuperscript{2} ASE = Average Standard Error}\\
\multicolumn{6}{l}{\rule{0pt}{1em}\textsuperscript{3} RMSE = Root Mean Square Error}\\
\addlinespace[0.5em]
\end{tabular}
\end{table}

\section{Illustration on real-data examples}
\label{s:application}
For illustration, post-hoc efficacy analyses were performed on three randomized clinical trials ($R1$, $R2_{loc}$, and $R2_{pulm}$) involving patients with Ewing Sarcoma to evaluate different consolidation treatments. After receiving intensive induction chemotherapy and surgery, patients were included in one of these trials according to prognostic factors and the response after surgery. In all these trials, the main endpoint was the event-free survival (EFS), defined as the time from random assignment to the first occurrence of any of the following events: relapse, second malignancy, or death from any cause, and the secondary endpoint was overall survival (OS), considering all causes of death.

The $R1$ trial was a phase III non-inferiority trial, which included standard-risk patients with small localized tumors or good histologic response to chemotherapy. The cyclophosphamide-based experimental arm was compared to the Ifosfamide-based control arm \citep{EWINGR1}. This trial recruited $856$ patients ($n=431$ received Vincristine-Actinomycine-Cyclophosphamide (VAC), and $n=425$ received Vincristine-Actinomycine-Ifosfamide (VAI)). The median follow-up was $5.9$ years, and the censoring rate was $73\%$ for the main endpoint. 
The $R2_{loc}$ trial, a phase III superiority trial, included high-risk patients with large localized tumors or poor histologic response. Busulfan and Melphalan (BuMel) were compared with the standard chemotherapy VAI \citep{EWINGR2_loc}. This trial recruited $240$ patients ($n=122$ received BuMel and $n=118$ received VAI). The median follow-up was $7.8$ years, and the censoring rate was $56\%$ for the main endpoint. 
The $R2_{pulm}$ trial, a phase III superiority trial, enrolled patients with only pulmonary or plural metastases and compared VAI + BuMel with VAI + pulmonary radiotherapy (RT) \citep{EWINGR2_pulm}. This trial included $287$ patients ($n=144$ receiving VAI+BuMel and $n=134$ received VAI+RT). The median follow-up was $8.1$ years, and the censoring rate was $50\%$ for the main endpoint.

The same methods and settings from the simulation study were used to analyze the EWING data. Supporting information Figure S1, shows the Kaplan-Meier curves for the three trials and the corresponding pseudo-observations profiles for all patients. Focusing on the $R_1$ trial, most of the events were observed between 0 and 3 years post-randomization. Consequently, the last time point to compute pseudo-observations is at $2.81$ year. As most individuals are censored, we observed most of the pseudo-observations above $1$ or between $0$ and $1$. Similar observations can be drawn from the $R2_{loc}$ and $R2_{pulm}$ trials.

\begin{figure}[!ht]\centering
\includegraphics[width=16.5cm]{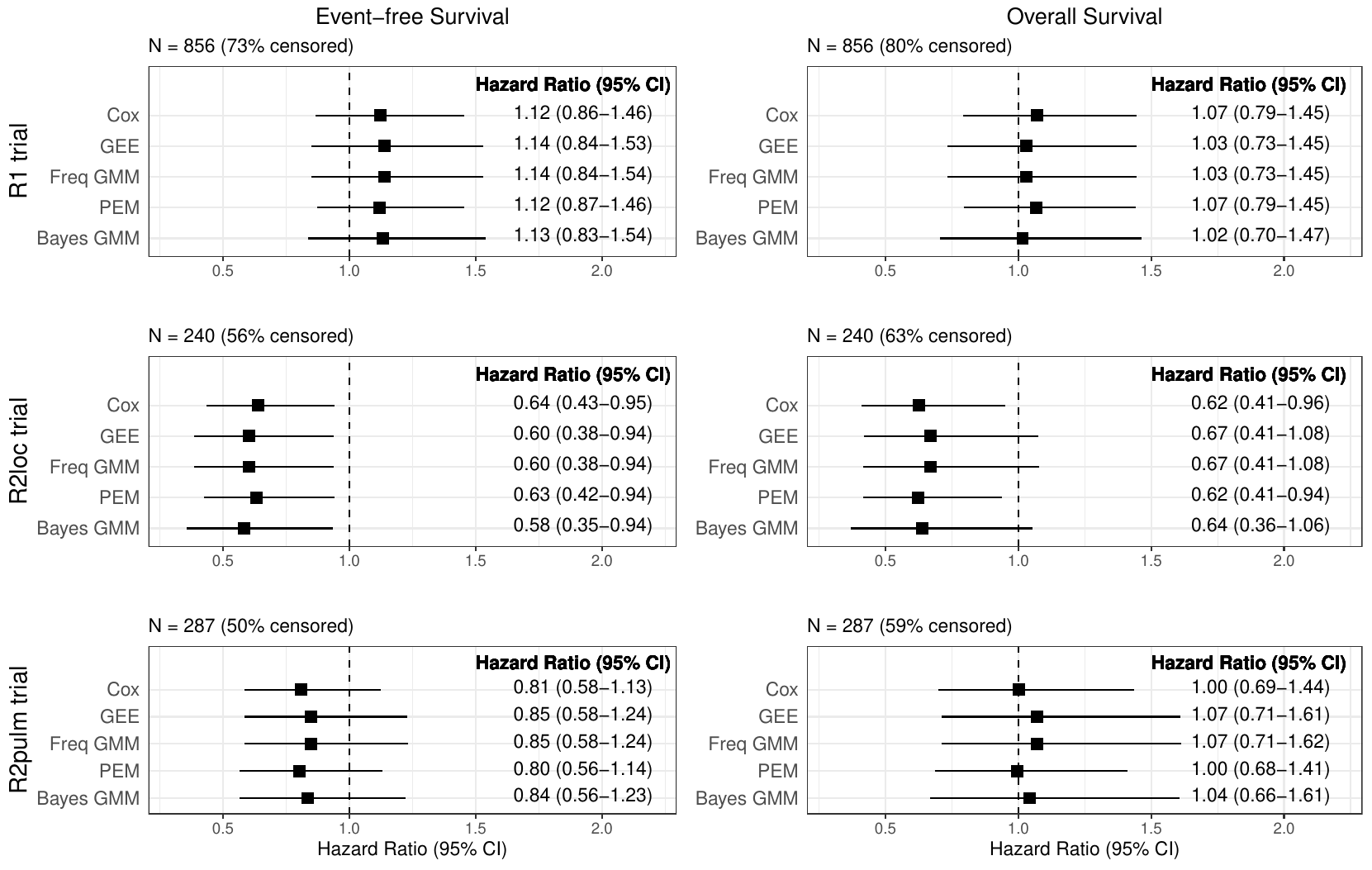}
\caption{Hazard ratio estimates (and 95\% confidence intervals) of treatment effect from the Cox proportional hazard, GEE, frequentist (Freq) GMM, piecewise exponential and Bayesian (Bayes) GMM models in the three EWING trials ($R1$, $R2_{loc}$, or $R2_{pulm}$) for event-free survival (left part) and overall survival (right part). These analyses were performed with no adjustment on covariates. The independent working correlation matrix is used for GEE and GMM approaches. The vertical dashed line represents the null effect.}
\label{orsini:fig3}
\end{figure}

Figure~\ref{orsini:fig3} depicts the estimates of the hazard ratios with their 95\% confidence intervals for EFS and OS, produced by the frequentist and Bayesian GMM and the three benchmark methods (Cox, GEE, and piecewise exponential models) without adjustment on covariates contrary to the published results. The results of the different methods are consistent, supporting the validity of the GMM approaches for analyzing pseudo-observations. Frequentist GMM and GEE give similar results with a higher variance, as expected, compared to the Cox proportional hazard model. The results from the Bayesian GMM and piecewise exponential model are also similar, with a slightly higher variance for the former. According to the trace plots of the NUTS sampling for the Bayesian GMM, the $3$ chains mixed well and appeared stationary, suggesting no divergence issue (see supporting information Figures S2a to S2f).

While the previous results are obtained with an independent working correlation matrix for GEE and GMM approaches (which is used by default for pseudo-observations analysis), our analysis was firstly extended using different correlation assumptions, as the generalized methods of moments also allow the definition of complex working correlation structures using multiple base matrices. We implemented the exchangeable (EXCH) and the first-order auto-regressive (AR-1) correlation matrices so that the results remain comparable with the GEE approach. However, the choice is not limited to these cases. Only one of the three chains did not converge for the Bayesian GMM with an exchangeable matrix for EFS. This issue was resolved by changing $\epsilon$ from 0.05 to 0.03 for generating initial values of the NUTS algorithm. Overall, the results using different working correlation matrices are similar within each approach and between approaches, except for the exchangeable matrix for situations where the treatment effect $\beta_1$ is close to 0 (log scale). The results reported in Table~\ref{orsini:tab5} for $R1$ trial (supporting information Tables S5a and S5b for $R2_{loc}$ and $R2_{pulm}$) suggest that increasing the complexity of the working correlation matrix gives a negligible increase in precision.

To further illustrate the versatility of our approach, the analysis of the three trials was secondly extended by including age as a covariate. Age was a stratification variable, reported as a binary variable ($<25$ or $\geq 25$) years in the $R1$ trial and as a four-categorical variable ($<12$, $12-18$, $18-25$, and $>25$) years in $R2_{loc}$ and $R2_{pulm}$ trials. The age-adjusted treatment effect is similar across the methods within a trial, with a larger variance for the Bayesian GMM in the $R2_{pulm}$ trial (supporting information Figure S3). One relevant advantage of the Bayesian inference over the frequentist approach is that one can better characterize the treatment effect through its posterior distribution, estimated with the piecewise exponential model and the Bayesian GMM based on pseudo-observations (supporting information Figure S4 for EFS endpoint). For example, in $R1$ trial, the posterior probability of the log HR to be below the noninferiority margin ($\log( 1.43)$) is $0.973$ (se=$0.002$) and $0.906$ (se=$0.005$) for piecewise exponential and GMM models, respectively. In $R2_{loc}$, the posterior probability of a log HR to be below the log HR under $H_1$ ($\log(0.60)$) is $0.365$ (se=$0.005$) and $0.552$ (se=$0.009$) for piecewise exponential and GMM models, respectively. In $R2_{pulm}$, the posterior probability of a log HR to be below the log HR under $H_1$ ($\log(0.65)$) is $0.091$ (se=$0.003$) and $0.148$ (se=$0.006$) for piecewise exponential and GMM models, respectively.

\begin{table}[!ht]\centering
\caption{Log hazard ratio and standard error of treatment effect estimated by GEE and GMM with different correlation matrices: Independence (IND),  Exchangeable (EXCH) and first-order auto-regressive (AR-1) in R1 trial for event-free survival and overall survival}
\label{orsini:tab5}
\medskip
\begin{tabular}{llll}
\toprule
Methods & WCM$^1$ & log(HR) & SE$^2$\\
\midrule
\addlinespace[0.3em]
\multicolumn{4}{l}{\textit{Event-free survival}}\\
\multicolumn{4}{l}{\textbf{Frequentist}}\\
\hspace{1em}GEE& IND & 0.1309 & 0.150\\
\hspace{1em}GEE & EXCH & 0.1015 & 0.148\\
\hspace{1em}GEE& AR-1 & 0.1268 & 0.148\\
\hspace{1em}GMM& IND & 0.1309 & 0.150\\
\hspace{1em}GMM & EXCH & 0.0819 & 0.149\\
\hspace{1em}GMM& AR-1 & 0.1223 & 0.146\\
\addlinespace[0.3em]
\multicolumn{4}{l}{\textbf{Bayesian}}\\
\hspace{1em}GMM& IND & 0.1253 & 0.156\\
\hspace{1em}GMM & EXCH & 0.0758 & 0.140\\
\hspace{1em}GMM& AR-1 & 0.1200 & 0.156\\
\midrule
\addlinespace[0.3em]
\multicolumn{4}{l}{\textit{Overall survival}}\\
\multicolumn{4}{l}{\textbf{Frequentist}}\\
\hspace{1em}GEE & IND & 0.0291 & 0.173\\
\hspace{1em}GEE & EXCH & 0.0164 & 0.170\\
\hspace{1em}GEE & AR-1 & 0.0251 & 0.171\\
\hspace{1em}GMM & IND & 0.0291 & 0.173\\
\hspace{1em}GMM & EXCH & 0.0068 & 0.171\\
\hspace{1em}GMM & AR-1 & 0.0110 & 0.168\\
\addlinespace[0.3em]
\multicolumn{4}{l}{\textbf{Bayesian}}\\
\hspace{1em}GMM & IND & 0.0155 & 0.184\\
\hspace{1em}GMM & EXCH & -0.0124 & 0.181\\
\hspace{1em}GMM & AR-1 & 0.0144 & 0.184\\
\bottomrule
\multicolumn{4}{l}{\rule{0pt}{1em}\textsuperscript{1} WCM = Working Correlation Matrix}\\
\multicolumn{4}{l}{\rule{0pt}{1em}\textsuperscript{2} SE = Standard Error}\\
\addlinespace[0.5em]
\end{tabular}
\end{table}

\section{Discussion}
\label{s:discuss}
In this paper, we propose a new and practical approach for Bayesian survival regression modeling based on pseudo-observations. This method does not require specifying the full likelihood, contrary to the usual Bayesian parametric and semi-parametric regression models, where nuisance parameters are specified to model the baseline hazard function. With this new approach, we bypass this specification by transforming time-to-event data into pseudo-observations, then analyzed by the generalized method of moments. The generalized method of moments applied to pseudo-observations was evaluated for estimating the hazard ratio from a two-arm randomized clinical trial in a frequentist and Bayesian framework. This approach results in valid inferences and comparable results to those produced by the Cox, GEE, and piecewise exponential models. In the frequentist framework, GMM and GEE have similar results regardless of the treatment effects, censoring rates, and sample sizes. More interestingly, in the Bayesian framework, the GMM gives unbiased results when a reasonable number of events is observed. The hazard ratio estimations are, as expected, less efficient compared to the piecewise exponential model, but remain acceptable given that the model does not require any assumption on the full likelihood contrary to the piecewise exponential model. The effect size of the treatment did not influence these results, and there was no inflation of the standard errors estimated by the Bayesian GMM compared to the frequentist GMM. Under the exchangeable or the first-order auto-regressive assumption, the GMM approach gave results similar to the GEE's. 

Although our approach makes valid inferences, special care is needed in setting adequate priors and starting values of the regression coefficients to avoid convergence problems during the sampling. These convergence issues occur when one value of a regression coefficient falls outside the support of the pseudo-likelihood function. Priors have to be appropriately chosen according to a range of reasonable values that can be taken by the regression coefficient. In our simulation study of two-arm randomized clinical trials, we built a procedure as a preliminary step to specify the initial values of the regression coefficients carefully.  So, both adequate priors and well-chosen initial values ensure the robustness of the Bayesian results. Although no additional covariate was included in the simulation study, the possibility of performing Bayesian inference with multiple covariates was reported in the real-data applications. As our approach was based on a pseudo-likelihood, it does not allow us to make predictions, while Bayesian deep learning algorithms using pseudo-observations have been proposed for this purpose \citep{Zhao2020}.

In conclusion, this paper proposes the first Bayesian modeling of pseudo-observations using the generalized method of moments, combining the advantages of pseudo-observations with those of Bayesian inference. The Bayesian generalized method of moments was based on pseudo-observations to estimate hazard ratios, one of the most straightforward applications of pseudo-observations. Although this application is not complex and classical Bayesian survival models may also be used, the aim of this work is to be a proof of concept that pseudo-observations can be analyzed in the Bayesian framework. It serves as a starting point before its extension to other applications where pseudo-observations are useful, such as the restricted mean survival time estimation, the estimation of transition and state probabilities derived from multi-state models, or the analysis for interval-censored data in a Bayesian framework, and unlocks various options not available with traditional proportional hazard models. Compared to the frequentist methods, this approach offers not only new insights on interpretation via the estimated posterior distribution but may also overcome their limitations in some situations, especially in randomized clinical trials for rare diseases where it allows enriching the analysis by incorporating external data.

\section*{Additional information}
The R code to analyze pseudo-observations using frequentist and Bayesian GMM is available on the Oncostat team's GitHub https://github.com/Oncostat/pseudo\_gmm

 \section*{Conflict of interest}

 The authors declare that they have no conflict of interest.

 \section*{Funding}
 This study was funded by PhD grant MESRI from the doctoral School of Public Health, Paris-Saclay University. 

%\bibliography{LDA}   % name your BibTeX data base

\begin{thebibliography}{}
        \bibitem[\protect \citeauthoryear{\nobreak Andersen, \nobreak Klein, and Rosthøj}{Andersen et~al.}{2003}]{Andersen2003}
        Andersen P. K., Klein J. P., and Rosthøj S. (2003) Generalised linear models for correlated pseudo‐observations, with applications to multi‐state models. 
          {\it Biometrika}, 90(1), 15–27.
        \bibitem[\protect \citeauthoryear{Andersen and Pohar-Perme}{2010}]{Andersen2010}
        Andersen, P. K., and Pohar-Perme, M. (2010) Pseudo-observations in survival analysis.
        {\it Statistical Methods in Medical Research}, 19(1), 71–99.
        \bibitem[\protect \citeauthoryear{Biard et al.}{2021}]{Biard2021}
        Biard, L., Bergeron, A., Lévy, V., and Chevret, S. (2021) Bayesian survival analysis for early detection of treatment effects in phase 3 clinical trials.
          {\it Contemporary Clinical Trials Communications}, 21, 100709.
        \bibitem[\protect \citeauthoryear{Bouaziz}{2023}]{Bouaziz2023}
        Bouaziz, O. (2023) Fast approximations of pseudo-observations in the context of right censoring and interval censoring.
          {\it Biometrical Journal}, 65(4), 2200071.
        \bibitem[\protect \citeauthoryear{Brard et al.}{2017}]{Brard2017}
        Brard, C., Le Teuff, G., Le Deley, M.-C., and Hampson, L. V. (2017) Bayesian survival analysis in clinical trials: What methods are used in practice?
          {\it Clinical Trials}, 14(1), 78–87.
        \bibitem[\protect \citeauthoryear{Carpenter et al.}{2017}]{Carpenter2017}
        Carpenter, B., Gelman, A., Hoffman, M. D., Lee, D., Goodrich, B., Betancourt, M., et al. (2017) Stan: A Probabilistic Programming Language. {\it Journal of Statistical Software}, 76(1), 1-32.
        \bibitem[\protect \citeauthoryear{Chernozhukov and Hong}{2003}]{Chernozhukov2003}
        Chernozhukov V., and Hong, H. (2003) An MCMC approach to classical estimation. {\it Journal of Econometrics}, 115(2), 293-346.
        \bibitem[\protect \citeauthoryear{Chevret}{2011}]{Chevret2011}
        Chevret, S. (2012). Bayesian adaptive clinical trials: A dream for statisticians only?
          {\it Statistics in Medicine}, 31(11–12), 1002–1013.
        \bibitem[\protect \citeauthoryear{Cox}{1972}]{Cox1972}
        Cox, D. R. (1972). Regression Models and Life-Tables. {\it Journal of the Royal Statistical Society: Series B (Methodological)}, 34(2), 187–202.
        \bibitem[\protect \citeauthoryear{Dirksen et al.}{2019}]{EWINGR2_pulm} 
        Dirksen, U., Brennan, B., Le Deley, M.-C., Cozic, N., van den Berg, H., Bhadri, V., et al. (2019). High-Dose Chemotherapy Compared With Standard Chemotherapy and Lung Radiation in Ewing Sarcoma With Pulmonary Metastases: Results of the European Ewing Tumour Working Initiative of National Groups, 99 Trial and EWING 2008.
        {\it Journal of Clinical Oncology}, 37(34), 3192–3202. 
        \bibitem[\protect \citeauthoryear{Fors and González}{2020}]{Fors2020}
        Fors, M., and González, P. (2020) Current status of Bayesian clinical trials for oncology. {\it Contemporary Clinical Trials Communications}, 20, 100658.
        \bibitem[\protect \citeauthoryear{Gelman et al.}{2008}]{Gelman2008}
        Gelman, A., Jakulin, A., Pittau, M. G., and Su, Y.-S. (2008) A weakly informative default prior distribution for logistic and other regression models. {\it The Annals of Applied Statistics}, 2(4), 1360-1383.
        \bibitem[\protect \citeauthoryear{\nobreak Gelman, \nobreak Simpson, and Betancourt}{Gelman et al.}{2017}]{Gelman2017}
        Gelman, A., Simpson, D., and Betancourt, M. (2017) The Prior Can Often Only Be Understood in the Context of the Likelihood. {\it Entropy}, 19(10), 555.
        \bibitem[\protect \citeauthoryear{\nobreak Graw, \nobreak Gerds, and Schumacher}{Graw et al.}{2009}]{Graw2009}
        Graw, F., Gerds, T. A., and Schumacher, M. (2009) On pseudo-values for regression analysis in competing risks models. {\it Lifetime Data Analysis}, 15(2), 241–255.
        \bibitem[\protect \citeauthoryear{Hansen}{1982}]{Hansen1982}
        Hansen, L. P. (1982) Large Sample Properties of Generalized Method of Moments Estimators. {\it Econometrica}, 50(4), 1029.
        \bibitem[\protect \citeauthoryear{Held}{2020}]{Held2020} Book chapter Held, L. (2020). Bayesian Tail Probabilities for Decision Making. Lesaffre, E., Baio, G., and Boulanger, B. (Eds.). (2020). {\it Bayesian Methods in Pharmaceutical Research.}. Chapman and Hall/CRC.
        \bibitem[\protect \citeauthoryear{Hoffman and Gelman}{2014}]{Hoffman2014}
        Hoﬀman, M. D., and Gelman, A. (2014) The No-U-Turn Sampler: Adaptively Setting Path Lengths in Hamiltonian Monte Carlo. {\it Journal of Machine Learning Research}, 15(1), 1351–1381.
        \bibitem[\protect \citeauthoryear{Højsgaard et al.}{2022}]{Højsgaard2022} 
        Højsgaard, S., Halekoh, U., Yan, J., and Ekstrøm, C. T. (2022) {\it geepack} R package version 1.3.9, https://cran.r-project.org/web/packages/geepack/ (accessed on October 2022).
        \bibitem[\protect \citeauthoryear{\nobreak Ibrahim, \nobreak Chen, and Sinha}{Ibrahim et al.}{2001}]{Ibrahim2001} 
        Ibrahim, J. G., Chen, M.-H., and Sinha, D. (2001) 
        {\it Bayesian survival analysis.}. Springer.
        \bibitem[\protect \citeauthoryear{Jacobsen and Martinussen}{2016}]{Jacobsen2016}
        Jacobsen, M., and Martinussen, T. (2016) A Note on the Large Sample Properties of Estimators Based on Generalized Linear Models for Correlated Pseudo‐observations. {\it Scandinavian Journal of Statistics}, 43(3), 845–862.
        \bibitem[\protect \citeauthoryear{\nobreak Jiang, \nobreak Song, and Kleinsasser}{Jiang et al.}{2019}]{Jiang2019} Jiang, Z., Song, P., and Kleinsasser, M. (2019) {\it qif: Quadratic Inference Function} R package version 1.5,   https://CRAN.R-project.org/package=qif (Accessed on April 2023)
        \bibitem[\protect \citeauthoryear{Kalbfleisch}{1978}]{Kalbfleisch1978} 
        Kalbfleisch, J. D. (1978). Non-Parametric Bayesian Analysis of Survival Time Data. {\it Journal of the Royal Statistical Society. Series B (Methodological)}, 40(2), 214–221.
        \bibitem[\protect \citeauthoryear{Klein and Andersen}{2005}]{Klein2005} 
        Klein, J. P., and Andersen, P. K. (2005). Regression modeling of competing risks data based on pseudovalues of the cumulative incidence function. {\it Biometrics}, 61(1), 223–229.
        \bibitem[\protect \citeauthoryear{Klein et al.}{2007}]{Klein2007} 
        Klein, J. P., Logan, B., Harhoff, M., and Andersen, P. K. (2007). Analyzing survival curves at a fixed point in time. {\it Statistics in Medicine}, 26(24), 4505–4519.
        \bibitem[\protect \citeauthoryear{Klein et al.}{2008}]{Klein2008} Klein, J. P., Gerster, M., Andersen, P. K., Tarima, S., and Perme, M. P. (2008) SAS and R functions to compute pseudo-values for censored data regression. {\it Computer Methods and Programs in Biomedicine}, 89(3), 289\,--\,300.
        \bibitem[\protect \citeauthoryear{Klein et al.}{2014}]{Klein2014} Klein, J. P., van Houwelingen, H. C., Ibrahim, J. G., and Scheike, T. H. (2014) {\it Handbook of survival analysis} CRC Press, Taylor and Francis Group.
        \bibitem[\protect \citeauthoryear{Le Deley et al.}{2014}]{EWINGR1} 
        Le Deley, M.-C., Paulussen, M., Lewis, I., Brennan, B., Ranft, A., Whelan, J., et al. (2014).Cyclophosphamide compared with ifosfamide in consolidation treatment of standard-risk Ewing sarcoma: Results of the randomized noninferiority Euro-EWING99-R1 trial.
        {\it Journal of the American Society of Clinical Oncology}, 32(23), 2440–2448. 
        \bibitem[\protect \citeauthoryear{Lesaffre et al.}{2024}]{Lesaffre2024} Lesaffre, E., Qi, H., Banbeta, A., and Rosmalen, J. van. (2024). A review of dynamic borrowing methods with applications in pharmaceutical research. {\it Brazilian Journal of Probability and Statistics}, 38(1), 1\,--\,31.
        \bibitem[\protect \citeauthoryear{Liang and Zeger}{1986}]{Liang1986} Liang, K.-Y. and Zeger, S. (1986) Longitudinal data analysis using generalized linear models. {\it Biometrika}, 73(1), 13\,--\,22.
        \bibitem[\protect \citeauthoryear{McCullagh and Nelder}{1991}]{McCullagh1991} McCullagh, P., and Nelder, J. A. (1991) {\it Generalized Linear Models (2nd ed)}. London: Chapman \& Hall.
        \bibitem[\protect \citeauthoryear{Murray et al.}{2014}]{Murray2014} Murray, T. A., Hobbs, B. P., Lystig, T. C., and Carlin, B. P. (2014). Semiparametric Bayesian commensurate survival model for post-market medical device surveillance with non-exchangeable historical data: Semiparametric Bayesian Commensurate Survival Model. 
        {\it Biometrics}, 70(1), 185–191.
        \bibitem[\protect \citeauthoryear{Murray et al.}{2016}]{Murray2016} 
        Murray, T. A., Hobbs, B. P., Sargent, D. J., and Carlin, B. P. (2016). Flexible Bayesian survival modeling with semiparametric time-dependent and shape-restricted covariate effects. 
        {\it Bayesian Analysis}, 11(2), 381-402.
        \bibitem[\protect \citeauthoryear{\nobreak Overgaard, \nobreak Parner, and Pedersen}{Overgaard et al.}{2017}]{Overgaard2017} 
        Overgaard, M., Parner, E. T., and Pedersen, J. (2017). Asymptotic theory of generalized estimating equations based on jack-knife pseudo-observations.
        {\it The Annals of Statistics}, 45(5), 1988–2015.
        \bibitem[\protect \citeauthoryear{\nobreak Pohar-Perme, \nobreak Gerster, and Rodrigues}{Pohar-Perme et al.}{2017}]{PoharPerme2017} 
        Pohar-Perme, M., Gerster, M., and Rodrigues, K. (2017) {\it pseudo: Computes Pseudo-Observations for Modeling} R package version 1.4.3 URL: https://cran.r-project.org/web/packages/pseudo/
        \bibitem[\protect \citeauthoryear{Qu et al.}{2000}]{Qu2000} 
        Qu, A., Lindsay, B. G., and Li, B. (2000) Improving generalised estimating equations using quadratic inference functions. {\it Biometrika}, 87(4), 823\,--\,836.
        \bibitem[\protect \citeauthoryear{R Core Team}{2021}]{RCoreTeam2021} 
         R Core Team (2021). R: A language and environment for statistical computing. R Foundation for Statistical Computing, Vienna, Austria. URL https://www.R-project.org/.
        \bibitem[\protect \citeauthoryear{Sachs and Gabriel}{2022}]{Sachs2022}
        Sachs, M. C., and Gabriel, E. E. (2022) Event History Regression with Pseudo-Observations: Computational Approaches and an Implementation in R.
          {\it Journal of Statistical Software}, 102(9), 1–34.          
        \bibitem[\protect \citeauthoryear{Stan Development Team}{2020}]{StanDev2020} 
        Stan Development Team. (2020)  Prior Choice Recommendations. Available online https://github.com/stan-dev/stan/wiki/Prior-Choice-Recommendations (accessed on April 2023).
        \bibitem[\protect \citeauthoryear{Stan Development Team}{2023}]{StanDev2023} Stan Development Team (2023). RStan: the R interface to Stan, R package version 2.21.7, https://mc-stan.org/ (accessed on October 2022).
        \bibitem[\protect \citeauthoryear{Vehtari}{2021}]{Vehtari2021} 
        Vehtari, A., Gelman, A., Simpson, D., Carpenter, B., and Bürkner, P.-C. (2021). Rank-Normalization, Folding, and Localization: An Improved Rˆ for Assessing Convergence of MCMC (with Discussion). {\it Bayesian Analysis}, 16(2), 667–718.
        \bibitem[\protect \citeauthoryear{Wan}{2017}]{Wan2017} 
        Wan, F. (2017) Simulating survival data with predefined censoring rates for proportional hazards models. {\it Statistics in Medicine}, 36(5), 838–854.
        \bibitem[\protect \citeauthoryear{Whelan et al.}{2018}]{EWINGR2_loc} 
        Whelan, J., Le Deley, M.-C., Dirksen, U., Le Teuff, G., Brennan, B., Gaspar, N., et al. (2018). High-dose chemotherapy and blood autologous stem-cell rescue compared With standard chemotherapy in localized high-risk Ewing sarcoma: results of Euro-E.W.I.N.G.99 and Ewing-2008.
        {\it Journal of Clinical Oncology}, 36(31). 
        \bibitem[\protect \citeauthoryear{Yan and Fine}{2004}]{Yan2004} Yan, J., and Fine, J. (2004). Estimating equations for association structures. 
        {\it Statistics in Medicine}, 23(6), 859–874.
         \bibitem[\protect \citeauthoryear{Yin}{2009}]{Yin2009} Yin, G. (2009) Bayesian generalized method of moments. {\it Bayesian Analysis}, 4(2), 191\,--\,208.
        \bibitem[\protect \citeauthoryear{\nobreak Yu, \nobreak Li, and Turner}{Yu et al.}{2020}]{Yu2020} 
        Yu, H., Li, F., and Turner, E. L. (2020). An evaluation of quadratic inference functions for estimating intervention effects in cluster randomized trials. {\it Contemporary Clinical Trials Communications}, 19, 100605.   
        \bibitem[\protect \citeauthoryear{Zhao and Feng}{2020}]{Zhao2020} Zhao, L., and Feng, D. (2020). Deep neural networks for survival analysis using pseudo values. 
        {\it IEEE Journal of Biomedical and Health Informatics}, 24(11), 3308–3314.  
        \bibitem[\protect \citeauthoryear{\nobreak Zhou, \nobreak Hanson, and Zhang}{Zhou et al.}{2020}]{Zhou2020} Zhou, H., Hanson, T., and Zhang J. (2020) spBayesSurv: Fitting Bayesian spatial survival models using R. 
        {\it Journal of Statistical Software}, 92(9), 1-33.  
        \bibitem[\protect \citeauthoryear{Ziegler}{1995}]{Ziegler1995} 
        Ziegler, A. (1995) The different parameterizations of the GEE1 and the GEE2.
        {\it Lecture Notes in Statistics}, 104, 315–324. 
        \end{thebibliography}
%\printbibliography


\end{document}

\maketitle

\begin{table}\centering
\captionsetup{labelformat=empty}
\caption{ \textbf{Table S1a:} Performances of the GMM models compared to Cox, GEE, and piecewise exponential (PEM) models with different sample sizes. The true log hazard ratio is fixed at $-0.1$ (HR=0.9) and the censoring rate at $20\%$.}
\medskip
\begin{tabular}{llrrrr}
\toprule
n&Methods&Bias&ASE$^1$&RMSE$^2$&Cov.$^3$\\ %
\midrule

\addlinespace[0.3em]
\textbf{50}&\multicolumn{5}{l}{\textbf{Frequentist}}\\
&\hspace{1em}Cox & -0.0141 & 0.325 & 0.329 & 94.8\\
&\hspace{1em}GEE & -0.0058 & 0.353 & 0.382 & 92.2\\
&\hspace{1em}GMM & -0.0058 & 0.365 & 0.382 & 93.4\\
\addlinespace[0.3em]
&\multicolumn{5}{l}{\textbf{Bayesian}}\\
&\hspace{1em}PEM & -0.0288 & 0.331 & 0.362 & 92.6\\
&\hspace{1em}GMM & -0.0927 & 0.350 & 0.388 & 91.1\\
\bottomrule

\addlinespace[0.3em]
\textbf{100}& \multicolumn{5}{l}{\textbf{Frequentist}}\\
&\hspace{1em}Cox & 0.0084 & 0.227 & 0.242 & 93.3\\
&\hspace{1em}GEE & 0.0134 & 0.252 & 0.269 & 93.5\\
&\hspace{1em}GMM & 0.0134 & 0.256 & 0.269 & 93.9\\
\addlinespace[0.3em]
&\multicolumn{5}{l}{\textbf{Bayesian}}\\
&\hspace{1em}PEM & -0.0002 & 0.232 & 0.261 & 91.2\\
&\hspace{1em}GMM & -0.0380 & 0.251 & 0.271 & 92.2\\
\bottomrule

\addlinespace[0.3em]
\textbf{200}&\multicolumn{5}{l}{\textbf{Frequentist}}\\
&\hspace{1em}Cox & 0.0004 & 0.159 & 0.159 & 94.0\\
&\hspace{1em}GEE & 0.0013 & 0.179 & 0.189 & 93.3\\
&\hspace{1em}GMM & 0.0013 & 0.180 & 0.189 & 93.7\\
\addlinespace[0.3em]
&\multicolumn{5}{l}{\textbf{Bayesian}}\\
&\hspace{1em}PEM & -0.0073 & 0.163 & 0.171 & 92.7\\
&\hspace{1em}GMM & -0.0064 & 0.187 & 0.196 & 92.8\\
\bottomrule

\addlinespace[0.3em]
\textbf{500}&\multicolumn{5}{l}{\textbf{Frequentist}}\\
&\hspace{1em}Cox & 0.0029 & 0.100 & 0.100 & 94.1\\
&\hspace{1em}GEE & 0.0041 & 0.114 & 0.112 & 94.9\\
&\hspace{1em}GMM & 0.0041 & 0.114 & 0.112 & 95.0\\
\addlinespace[0.3em]
&\multicolumn{5}{l}{\textbf{Bayesian}}\\
&\hspace{1em}PEM & -0.0006 & 0.101 & 0.103 & 94.0\\
&\hspace{1em}GMM & 0.0012 & 0.116 & 0.113 & 94.9\\
\bottomrule

\addlinespace[0.3em]
\textbf{1000}&\multicolumn{5}{l}{\textbf{Frequentist}}\\
&\hspace{1em}Cox & 0.0028 & 0.071 & 0.072 & 95.0\\
&\hspace{1em}GEE & 0.0014 & 0.080 & 0.080 & 95.5\\
&\hspace{1em}GMM & 0.0014 & 0.080 & 0.080 & 95.5\\
\addlinespace[0.3em]
&\multicolumn{5}{l}{\textbf{Bayesian}}\\
&\hspace{1em}PEM & 0.0011 & 0.071 & 0.073 & 94.4\\
&\hspace{1em}GMM & 0.0000 & 0.081 & 0.081 & 95.3\\
\bottomrule
\multicolumn{6}{l}{\rule{0pt}{1em}\textsuperscript{1} ASE = Average Standard Error}\\
\multicolumn{6}{l}{\rule{0pt}{1em}\textsuperscript{2} RMSE = Root Mean Square Error}\\
\multicolumn{6}{l}{\rule{0pt}{1em}\textsuperscript{3} Cov. = Coverage}\\
\addlinespace[0.5em]
\end{tabular}
\end{table}

\begin{table}\centering
\captionsetup{labelformat=empty}
\caption{ \textbf{Table S1b:} Performances of the GMM models compared to Cox, GEE, and piecewise exponential (PEM) models with different sample sizes. The true log hazard ratio is fixed at $-0.5$ (HR=0.6) and the censoring rate at $20\%$.}
\medskip
\begin{tabular}{llrrrr}
\toprule
n&Methods&Bias&ASE$^1$&RMSE$^2$&Cov.$^3$\\ %
\midrule

\addlinespace[0.3em]
\textbf{50}&\multicolumn{5}{l}{\textbf{Frequentist}}\\
&\hspace{1em}Cox & -0.0240 & 0.330 & 0.339 & 95.5\\
&\hspace{1em}GEE & -0.0200 & 0.359 & 0.396 & 92.4\\
&\hspace{1em}GMM & -0.0200 & 0.372 & 0.396 & 93.8\\
\addlinespace[0.3em]
&\multicolumn{5}{l}{\textbf{Bayesian}}\\
&\hspace{1em}PEM & -0.0802 & 0.335 & 0.376 & 92.0\\
&\hspace{1em}GMM & -0.0763 & 0.357 & 0.387 & 91.2\\
\bottomrule

\addlinespace[0.3em]
\textbf{100}& \multicolumn{5}{l}{\textbf{Frequentist}}\\
&\hspace{1em}Cox & 0.0026 & 0.230 & 0.250 & 93.7\\
&\hspace{1em}GEE & 0.0076 & 0.256 & 0.276 & 94.2\\
&\hspace{1em}GMM & 0.0076 & 0.260 & 0.276 & 94.4\\
\addlinespace[0.3em]
&\multicolumn{5}{l}{\textbf{Bayesian}}\\
&\hspace{1em}PEM & -0.0327 & 0.234 & 0.268 & 91.6\\
&\hspace{1em}GMM & -0.0295 & 0.256 & 0.275 & 92.5\\
\bottomrule

\addlinespace[0.3em]
\textbf{200}&\multicolumn{5}{l}{\textbf{Frequentist}}\\
&\hspace{1em}Cox & -0.0017 & 0.161 & 0.162 & 94.4\\
&\hspace{1em}GEE & 0.0002 & 0.182 & 0.190 & 92.9\\
&\hspace{1em}GMM & 0.0002 & 0.183 & 0.190 & 93.1\\
\addlinespace[0.3em]
&\multicolumn{5}{l}{\textbf{Bayesian}}\\
&\hspace{1em}PEM & -0.0329 & 0.164 & 0.176 & 93.0\\
&\hspace{1em}GMM & -0.0239 & 0.192 & 0.199 & 93.0\\
\bottomrule

\addlinespace[0.3em]
\textbf{500}& \multicolumn{5}{l}{\textbf{Frequentist}}\\
&\hspace{1em}Cox & 0.0034 & 0.102 & 0.101 & 94.8\\
&\hspace{1em}GEE & 0.0020 & 0.115 & 0.112 & 95.0\\
&\hspace{1em}GMM & 0.0020 & 0.115 & 0.112 & 95.1\\
\addlinespace[0.3em]
&\multicolumn{5}{l}{\textbf{Bayesian}}\\
&\hspace{1em}PEM & -0.0107 & 0.102 & 0.104 & 94.1\\
&\hspace{1em}GMM & -0.0071 & 0.118 & 0.114 & 95.6\\
\bottomrule

\addlinespace[0.3em]
\textbf{1000}&\multicolumn{5}{l}{\textbf{Frequentist}}\\
&\hspace{1em}Cox & 0.0034 & 0.072 & 0.072 & 95.2\\
&\hspace{1em}GEE & 0.0014 & 0.081 & 0.082 & 94.9\\
&\hspace{1em}GMM & 0.0014 & 0.082 & 0.082 & 94.9\\
\addlinespace[0.3em]
&\multicolumn{5}{l}{\textbf{Bayesian}}\\
&\hspace{1em}PEM & -0.0039 & 0.072 & 0.074 & 94.9\\
&\hspace{1em}GMM & -0.0030 & 0.082 & 0.083 & 95.0\\

\bottomrule
\multicolumn{6}{l}{\rule{0pt}{1em}\textsuperscript{1} ASE = Average Standard Error}\\
\multicolumn{6}{l}{\rule{0pt}{1em}\textsuperscript{2} RMSE = Root Mean Square Error}\\
\multicolumn{6}{l}{\rule{0pt}{1em}\textsuperscript{3} Cov. = Coverage}\\
\addlinespace[0.5em]
\end{tabular}
\end{table}

\begin{table}\centering
\captionsetup{labelformat=empty}
\caption{ \textbf{Table S2a:} Performances of the GMM models compared to Cox, GEE, and piecewise exponential (PEM) models with different censoring rates. The true log hazard ratio is fixed at $-0.1$ (HR=0.9) and the sample size at $500$.}
\medskip
\begin{tabular}{llrrrr}
\toprule
C.R.$^1$&Methods&Bias&ASE$^2$&RMSE$^3$&Cov.$^4$\\
\midrule

\addlinespace[0.3em]
\textbf{5\%}&\multicolumn{5}{l}{\textbf{Frequentist}}\\
&\hspace{1em}Cox & 0.0023 & 0.092 & 0.093 & 94.3\\
&\hspace{1em}GEE & 0.0029 & 0.106 & 0.105 & 95.5\\
&\hspace{1em}GMM & 0.0029 & 0.107 & 0.105 & 95.5\\
\addlinespace[0.3em]
&\multicolumn{5}{l}{\textbf{Bayesian}}\\
&\hspace{1em}PEM & -0.0012 & 0.093 & 0.097 & 93.9\\
&\hspace{1em}GMM & 0.0000 & 0.108 & 0.107 & 95.7\\
\bottomrule

\addlinespace[0.3em]
\textbf{10\%}& \multicolumn{5}{l}{\textbf{Frequentist}}\\
&\hspace{1em}Cox & 0.0015 & 0.095 & 0.094 & 94.8\\
&\hspace{1em}GEE & 0.0020 & 0.108 & 0.105 & 95.4\\
&\hspace{1em}GMM & 0.0020 & 0.109 & 0.105 & 95.4\\
\addlinespace[0.3em]
&\multicolumn{5}{l}{\textbf{Bayesian}}\\
&\hspace{1em}PEM & -0.0022 & 0.096 & 0.098 & 93.8\\
&\hspace{1em}GMM & -0.0076 & 0.108 & 0.106 & 95.2\\
\bottomrule

\addlinespace[0.3em]
\textbf{20\%}&\multicolumn{5}{l}{\textbf{Frequentist}}\\
&\hspace{1em}Cox & 0.0029 & 0.100 & 0.100 & 94.1\\
&\hspace{1em}GEE & 0.0041 & 0.114 & 0.112 & 94.9\\
&\hspace{1em}GMM & 0.0041 & 0.114 & 0.112 & 95.0\\
\addlinespace[0.3em]
&\multicolumn{5}{l}{\textbf{Bayesian}}\\
&\hspace{1em}PEM & -0.0006 & 0.101 & 0.103 & 94.0\\
&\hspace{1em}GMM & 0.0012 & 0.116 & 0.113 & 94.9\\
\bottomrule

\addlinespace[0.3em]
\textbf{30\%}& \multicolumn{5}{l}{\textbf{Frequentist}}\\
&\hspace{1em}Cox & 0.0050 & 0.107 & 0.107 & 95.4\\
&\hspace{1em}GEE & 0.0052 & 0.120 & 0.117 & 95.5\\
&\hspace{1em}GMM & 0.0052 & 0.121 & 0.117 & 95.7\\
\addlinespace[0.3em]
&\multicolumn{5}{l}{\textbf{Bayesian}}\\
&\hspace{1em}PEM & 0.0014 & 0.108 & 0.110 & 94.7\\
&\hspace{1em}GMM & 0.0023 & 0.123 & 0.119 & 95.4\\
\bottomrule

\addlinespace[0.3em]
\textbf{70\%}& \multicolumn{5}{l}{\textbf{Frequentist}}\\
&\hspace{1em}Cox & 0.0035 & 0.164 & 0.163 & 95.1\\
&\hspace{1em}GEE & 0.0018 & 0.183 & 0.184 & 94.9\\
&\hspace{1em}GMM & 0.0018 & 0.184 & 0.184 & 94.9\\
\addlinespace[0.3em]
&\multicolumn{5}{l}{\textbf{Bayesian}}\\
&\hspace{1em}PEM & 0.0008 & 0.165 & 0.169 & 94.2\\
&\hspace{1em}GMM & -0.0066 & 0.193 & 0.186 & 95.1\\
\bottomrule
\multicolumn{6}{l}{\rule{0pt}{1em}\textsuperscript{1} C.R. = Censoring Rate}\\
\multicolumn{6}{l}{\rule{0pt}{1em}\textsuperscript{2} ASE = Average Standard Error}\\
\multicolumn{6}{l}{\rule{0pt}{1em}\textsuperscript{3} RMSE = Root Mean Square Error}\\
\multicolumn{6}{l}{\rule{0pt}{1em}\textsuperscript{4} Cov. = Coverage}\\
\addlinespace[0.5em]
\end{tabular}
\end{table}

\begin{table}\centering
\captionsetup{labelformat=empty}
\caption{\textbf{Table S2b:} Performances of the GMM models compared to Cox, GEE, and piecewise exponential (PEM) models with different censoring rates. The true log hazard ratio is fixed at $-0.5$ (HR=0.6) and the sample size at $500$.}
\medskip
\begin{tabular}{llrrrr}
\toprule
C.R.$^1$&Methods&Bias&ASE$^2$&RMSE$^3$&Cov.$^4$\\
\midrule

\addlinespace[0.3em]
\textbf{5\%}&\multicolumn{5}{l}{\textbf{Frequentist}}\\
&\hspace{1em}Cox & 0.0015 & 0.094 & 0.096 & 94.2\\
&\hspace{1em}GEE & 0.0041 & 0.108 & 0.108 & 95.2\\
&\hspace{1em}GMM & 0.0041 & 0.108 & 0.108 & 95.4\\
\addlinespace[0.3em]
&\multicolumn{5}{l}{\textbf{Bayesian}}\\
&\hspace{1em}PEM & -0.0127 & 0.095 & 0.099 & 94.1\\
&\hspace{1em}GMM & -0.0062 & 0.111 & 0.110 & 94.9\\
\bottomrule

\addlinespace[0.3em]
\textbf{10\%}& \multicolumn{5}{l}{\textbf{Frequentist}}\\
&\hspace{1em}Cox & 0.0015 & 0.096 & 0.097 & 94.4\\
&\hspace{1em}GEE & 0.0035 & 0.110 & 0.108 & 95.2\\
&\hspace{1em}GMM & 0.0035 & 0.110 & 0.108 & 95.2\\
\addlinespace[0.3em]
&\multicolumn{5}{l}{\textbf{Bayesian}}\\
&\hspace{1em}PEM & -0.0121 & 0.097 & 0.100 & 94.7\\
&\hspace{1em}GMM & -0.0064 & 0.113 & 0.111 & 95.0\\
\bottomrule

\addlinespace[0.3em]
\textbf{20\%}&\multicolumn{5}{l}{\textbf{Frequentist}}\\
&\hspace{1em}Cox & 0.0034 & 0.102 & 0.101 & 94.8\\
&\hspace{1em}GEE & 0.0020 & 0.115 & 0.112 & 95.0\\
&\hspace{1em}GMM & 0.0020 & 0.115 & 0.112 & 95.1\\
\addlinespace[0.3em]
&\multicolumn{5}{l}{\textbf{Bayesian}}\\
&\hspace{1em}PEM & -0.0107 & 0.102 & 0.104 & 94.1\\
&\hspace{1em}GMM & -0.0071 & 0.118 & 0.114 & 95.6\\
\bottomrule

\addlinespace[0.3em]
\textbf{30\%}& \multicolumn{5}{l}{\textbf{Frequentist}}\\
&\hspace{1em}Cox & 0.0047 & 0.108 & 0.109 & 94.9\\
&\hspace{1em}GEE & 0.0016 & 0.122 & 0.121 & 94.9\\
&\hspace{1em}GMM & 0.0016 & 0.123 & 0.121 & 94.9\\
\addlinespace[0.3em]
&\multicolumn{5}{l}{\textbf{Bayesian}}\\
&\hspace{1em}PEM & -0.0091 & 0.109 & 0.112 & 94.2\\
&\hspace{1em}GMM & -0.0075 & 0.125 & 0.123 & 94.9\\
\bottomrule

\addlinespace[0.3em]
\textbf{70\%}& \multicolumn{5}{l}{\textbf{Frequentist}}\\
&\hspace{1em}Cox & 0.0020 & 0.167 & 0.165 & 94.5\\
&\hspace{1em}GEE & -0.0019 & 0.187 & 0.188 & 93.9\\
&\hspace{1em}GMM & -0.0019 & 0.188 & 0.188 & 94.1\\
\addlinespace[0.3em]
&\multicolumn{5}{l}{\textbf{Bayesian}}\\
&\hspace{1em}PEM & -0.0111 & 0.168 & 0.170 & 93.9\\
&\hspace{1em}GMM & -0.0185 & 0.198 & 0.191 & 94.6\\
\bottomrule
\multicolumn{6}{l}{\rule{0pt}{1em}\textsuperscript{1} C.R. = Censoring Rate}\\
\multicolumn{6}{l}{\rule{0pt}{1em}\textsuperscript{2} ASE = Average Standard Error}\\
\multicolumn{6}{l}{\rule{0pt}{1em}\textsuperscript{3} RMSE = Root Mean Square Error}\\
\multicolumn{6}{l}{\rule{0pt}{1em}\textsuperscript{4} Cov. = Coverage}\\
\addlinespace[0.5em]
\end{tabular}
\end{table}

\begin{table}\centering
\captionsetup{labelformat=empty}
\caption{ \textbf{Table S3:} Comparison of the performances of GEE and GMM models with different time points. The true log hazard ratio is fixed at $-0.3$ (HR=0.74), the censoring rate at $20\%$, and the sample size at $500$.} 
\medskip
\begin{tabular}{llrrrr}
\toprule
Methods & K & Bias & ASE$^1$ & RMSE$^2$ & Coverage\\
\midrule
\addlinespace[0.3em]
\multicolumn{6}{l}{\textbf{Frequentist}}\\
\hspace{1em}GEE & 5 & 0.0032 & 0.114 & 0.112 & 95.4\\
\hspace{1em} & 7 & 0.0024 & 0.112 & 0.110 & 95.3\\
\hspace{1em} & 10 & 0.0029 & 0.111 & 0.109 & 95.3\\
\hspace{1em}GMM & 5 & 0.0032 & 0.114 & 0.112 & 95.5\\
\hspace{1em} & 7 & 0.0024 & 0.113 & 0.110 & 95.3\\
\hspace{1em} & 10 & 0.0029 & 0.111 & 0.109 & 95.4\\
\addlinespace[0.3em]
\multicolumn{6}{l}{\textbf{Bayesian}}\\
\hspace{1em}GMM & 5 & -0.0028 & 0.116 & 0.113 & 95.4\\
\hspace{1em} & 7 & -0.0051 & 0.114 & 0.112 & 95.3\\
\hspace{1em} & 10 & -0.0063 & 0.113 & 0.112 & 95.0\\
\bottomrule
\multicolumn{6}{l}{\rule{0pt}{1em}\textsuperscript{1} AVE = Average Standard Error}\\
\multicolumn{6}{l}{\rule{0pt}{1em}\textsuperscript{2} RMSE = Root Mean Square Error}\\
\addlinespace[1em]
\end{tabular}
\end{table}

\begin{table}\centering
\captionsetup{labelformat=empty}
\caption{ \textbf{Table S4a:} Comparison of the performances of GEE and GMM models with different correlation matrices:  independence (IND), exchangeable (EXCH), and first-order auto-regressive (AR-1). The true log hazard ratio is fixed at $-0.1$ (HR=0.9), the censoring rate at $20\%$, and the sample size at $500$.}
\medskip
\begin{tabular}{llrrrr}
\toprule
Methods & WCM$^1$ & Bias & ASE$^2$ & RMSE$^3$ & Coverage\\
\midrule
\addlinespace[0.3em]
\multicolumn{6}{l}{\textbf{Frequentist}}\\
\hspace{1em}GEE & IND & 0.0041 & 0.114 & 0.112 & 94.9\\
\hspace{1em}GEE & EXCH & 0.0033 & 0.113 & 0.112 & 95.1\\
\hspace{1em}GEE & AR-1 & 0.0036 & 0.111 & 0.110 & 94.7\\
\hspace{1em}GMM & IND & 0.0041 & 0.114 & 0.112 & 95.0\\
\hspace{1em}GMM & EXCH & 0.0028 & 0.111 & 0.110 & 95.7\\
\hspace{1em}GMM & AR-1 & 0.0020 & 0.110 & 0.110 & 95.4\\
\addlinespace[0.3em]
\multicolumn{6}{l}{\textbf{Bayesian}}\\
\hspace{1em}GMM & IND & 0.0012 & 0.116 & 0.113 & 94.9\\
\hspace{1em}GMM & EXCH & -0.0002 & 0.112 & 0.111 & 95.2\\
\hspace{1em}GMM & AR-1 & -0.0009 & 0.113 & 0.112 & 94.6\\
\bottomrule
\multicolumn{6}{l}{\rule{0pt}{1em}\textsuperscript{1} WCM = Working Correlation Matrix}\\
\multicolumn{6}{l}{\rule{0pt}{1em}\textsuperscript{2} ASE = Average Standard Error}\\
\multicolumn{6}{l}{\rule{0pt}{1em}\textsuperscript{3} RMSE = Root Mean Square Error}\\
\addlinespace[0.5em]
\end{tabular}
\end{table}

\begin{table}\centering
\captionsetup{labelformat=empty}
\caption{\textbf{Table S4b:} Comparison of the performances of GEE and GMM models with different correlation matrices:  independence (IND), exchangeable (EXCH), and first-order auto-regressive (AR-1). The true log hazard ratio is fixed at $-0.5$ (HR=0.6), the censoring rate at $20\%$, and the sample size at $500$.}
\medskip
\begin{tabular}{llrrrr}
\toprule
Methods & WCM$^1$ & Bias & ASE$^2$ & RMSE$^3$ & Coverage\\
\midrule
\addlinespace[0.3em]
\multicolumn{6}{l}{\textbf{Frequentist}}\\
\hspace{1em}GEE & IND & 0.0020 & 0.115 & 0.112 & 95.0\\
\hspace{1em}GEE & EXCH & 0.0009 & 0.114 & 0.112 & 95.5\\
\hspace{1em}GEE & AR-1 & 0.0013 & 0.112 & 0.110 & 95.5\\
\hspace{1em}GMM & IND & 0.0020 & 0.115 & 0.112 & 95.1\\
\hspace{1em} GMM & EXCH & -0.0021 & 0.112 & 0.111 & 95.2\\
\hspace{1em} GMM & AR-1 & -0.0051 & 0.112 & 0.111 & 95.2\\
\addlinespace[0.3em]
\multicolumn{6}{l}{\textbf{Bayesian}}\\
\hspace{1em}GMM & IND & -0.0071 & 0.118 & 0.114 & 95.6\\
\hspace{1em}GMM & EXCH & -0.0105 & 0.114 & 0.113 & 95.2\\
\hspace{1em}GMM & AR-1 & -0.0138 & 0.114 & 0.114 & 94.8\\
\bottomrule
\multicolumn{6}{l}{\rule{0pt}{1em}\textsuperscript{1} WCM = Working Correlation Matrix}\\
\multicolumn{6}{l}{\rule{0pt}{1em}\textsuperscript{2} ASE = Average Standard Error}\\
\multicolumn{6}{l}{\rule{0pt}{1em}\textsuperscript{3} RMSE = Root Mean Square Error}\\
\addlinespace[0.5em]
\end{tabular}
\end{table}

\begin{table}\centering
\captionsetup{labelformat=empty}
\caption{ \textbf{Table S5a:} Log hazard ratio and standard error of the treatment effect estimated by GEE and GMM with different correlation matrices:  independence (IND), exchangeable (EXCH), and first-order auto-regressive (AR-1) in $R2_{loc}$ trial for event-free survival and overall survival}
\medskip
\begin{tabular}{llll}
\toprule
Methods & WCM$^1$ & log(HR) & SE$^2$\\
\midrule
\addlinespace[0.3em]
\multicolumn{4}{l}{\textit{Event-free survival}}\\
\multicolumn{4}{l}{\textbf{Frequentist}}\\
\hspace{1em}GEE & IND & -0.5073 & 0.226\\
\hspace{1em}GEE & EXCH & -0.5383 & 0.220\\
\hspace{1em}GEE & AR-1 & -0.4903 & 0.219\\
\hspace{1em}GMM& IND & -0.5073 & 0.227\\
\hspace{1em}GMM & EXCH & -0.5457 & 0.223\\
\hspace{1em}GMM& AR-1 & -0.4549 & 0.228\\
\addlinespace[0.3em]
\multicolumn{4}{l}{\textbf{Bayesian}}\\
\hspace{1em}GMM& IND & -0.5389 & 0.248\\
\hspace{1em}GMM & EXCH & -0.5143 & 0.241\\
\hspace{1em}GMM& AR-1 & -0.5025 & 0.281\\
\midrule
\addlinespace[0.3em]
\multicolumn{4}{l}{\textit{Overall survival}}\\
\multicolumn{4}{l}{\textbf{Frequentist}}\\
\hspace{1em}GEE & IND & -0.3995 & 0.240\\
\hspace{1em}GEE & EXCH & -0.4529 & 0.236\\
\hspace{1em}GEE & AR-1 & -0.4246 & 0.235\\
\hspace{1em}GMM & IND & -0.3995 & 0.242\\
\hspace{1em}GMM  & EXCH & -0.4736 & 0.241\\
\hspace{1em}GMM & AR-1 & -0.4755 & 0.247\\
\addlinespace[0.3em]
\multicolumn{4}{l}{\textbf{Bayesian}}\\
\hspace{1em}GMM& IND & -0.4507 & 0.270\\
\hspace{1em}GMM & EXCH & -0.5132 & 0.264\\
\hspace{1em}GMM& AR-1 & -0.5430 & 0.311\\
\bottomrule
\multicolumn{4}{l}{\rule{0pt}{1em}\textsuperscript{1} WCM = Working Correlation Matrix}\\
\multicolumn{4}{l}{\rule{0pt}{1em}\textsuperscript{2} SE = Standard Error}\\
\addlinespace[0.5em]
\end{tabular}
\end{table}

\begin{table}\centering
\captionsetup{labelformat=empty}
\caption{ \textbf{Table S5b:} Log hazard ratio and standard errors of the treatment effect estimated by GEE and GMM with different correlation matrices: independence (IND), exchangeable (EXCH), and first-order auto-regressive (AR-1) in $R2_{pulm}$ trial for event-free survival and overall survival}
\medskip
\begin{tabular}{llll}
\toprule
Methods & WCM$^1$ & log(HR) & SE$^2$\\
\midrule
\multicolumn{4}{l}{\textit{Event-free survival}}\\
\multicolumn{4}{l}{\textbf{Frequentist}}\\
\hspace{1em}GEE & IND & -0.1641 & 0.190\\
\hspace{1em}GEE  & EXCH & -0.2086 & 0.186\\
\hspace{1em}GEE & AR-1 & -0.1756 & 0.184\\
\hspace{1em}GMM& IND & -0.1641 & 0.191\\
\hspace{1em}GMM & EXCH & -0.2100 & 0.188\\
\hspace{1em}GMM& AR-1 & -0.1791 & 0.184\\
\addlinespace[0.3em]
\multicolumn{4}{l}{\textbf{Bayesian}}\\
\hspace{1em}GMM& IND & -0.1798 & 0.196\\
\hspace{1em}GMM & EXCH & -0.1956 & 0.197\\
\hspace{1em}GMM& AR-1 & -0.1985 & 0.199\\
\midrule
\addlinespace[0.3em]
\multicolumn{4}{l}{\textit{Overall survival}}\\
\multicolumn{4}{l}{\textbf{Frequentist}}\\
\hspace{1em}GEE & IND & 0.0676 & 0.208\\
\hspace{1em}GEE  & EXCH & 0.0095 & 0.204\\
\hspace{1em}GEE & AR-1 & 0.0469 & 0.203\\
\hspace{1em}GMM& IND & 0.0676 & 0.209\\
\hspace{1em}GMM & EXCH & 0.0029 & 0.208\\
\hspace{1em}GMM& AR-1 & 0.0240 & 0.203\\
\addlinespace[0.3em]
\multicolumn{4}{l}{\textbf{Bayesian}}\\
\hspace{1em}GMM& IND & 0.0424 & 0.221\\
\hspace{1em}GMM & EXCH & 0.0000 & 0.227\\
\hspace{1em}GMM& AR-1 & 0.0030 & 0.226\\
\bottomrule
\multicolumn{4}{l}{\rule{0pt}{1em}\textsuperscript{1} WCM = Working Correlation Matrix}\\
\multicolumn{4}{l}{\rule{0pt}{1em}\textsuperscript{2} SE = Standard Error}\\
\addlinespace[0.5em]
\end{tabular}
\end{table}
\newpage

\begin{figure}\centering
\includegraphics[width=15cm]{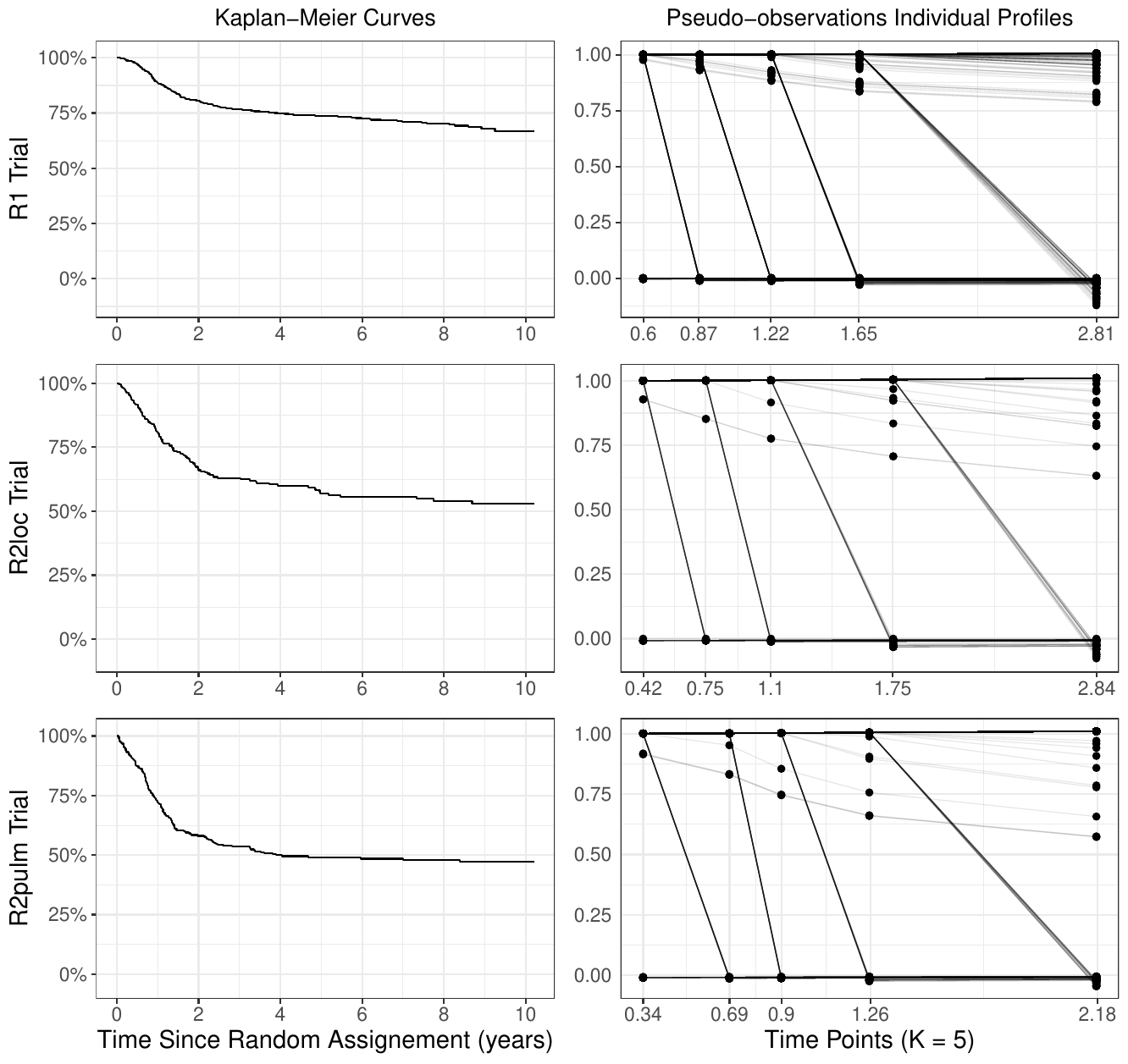}
\captionsetup{labelformat=empty}
\caption{ \textbf{Figure S1:} Kaplan-Meier curves of the event-free survival and corresponding pseudo-observations individual profiles for the EWING trials.}
\end{figure}

\begin{figure}\centering
\includegraphics[width=15cm]{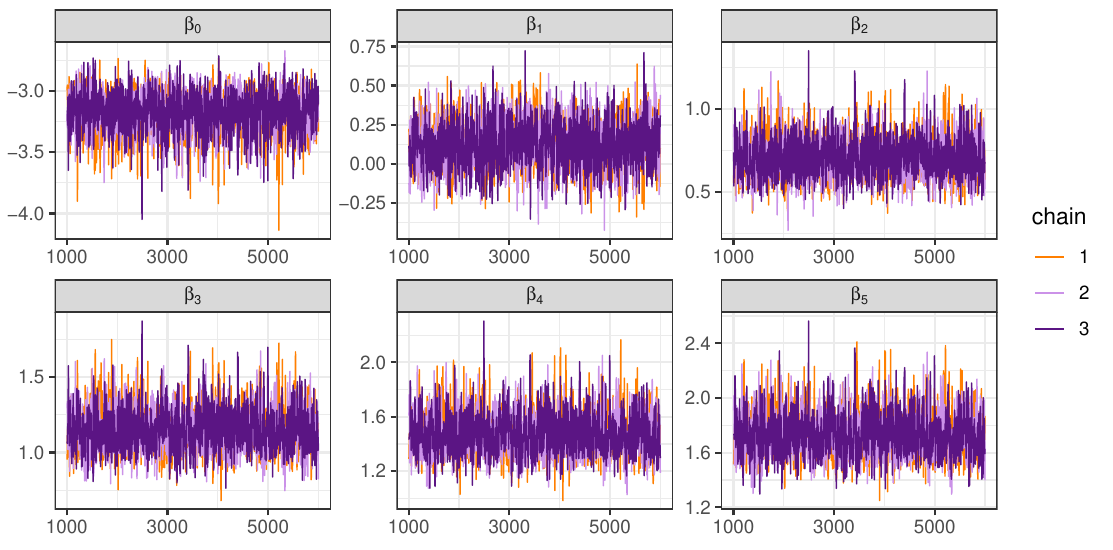}
\captionsetup{labelformat=empty}
\caption{ \textbf{Figure S2a:} Post warm-up MCMCs, using the Bayesian GMM to estimate HR based on the event-free survival in EWING (R1) trial. $\beta_0$ is the intercept, $\beta_1$ is the parameter of the treatment factor and $(\beta_2, ... \beta_5)$ are for the $K-1$ dummy time points variables.}
\end{figure}

\begin{figure}\centering
\includegraphics[width=15cm]{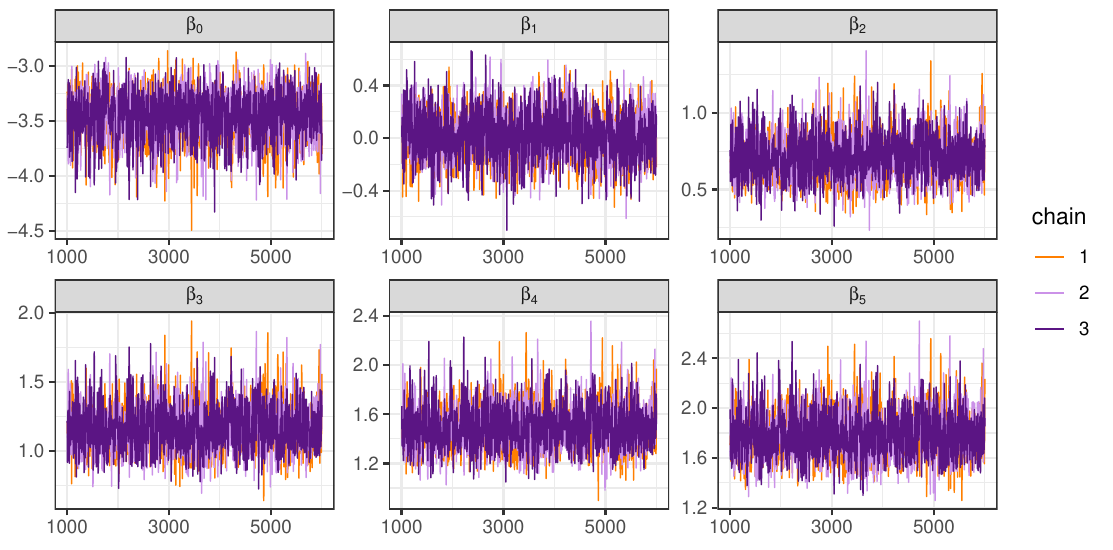}
\captionsetup{labelformat=empty}
\caption{ \textbf{Figure S2b:} Post warm-up MCMCs, using the Bayesian GMM to estimate HR based on the overall survival in EWING (R1) trial. $\beta_0$ is the intercept, $\beta_1$ is the parameter of the treatment factor and $(\beta_2, ... \beta_5)$ are for the $K-1$ dummy time points variables.}
\end{figure}

\begin{figure}\centering
\includegraphics[width=15cm]{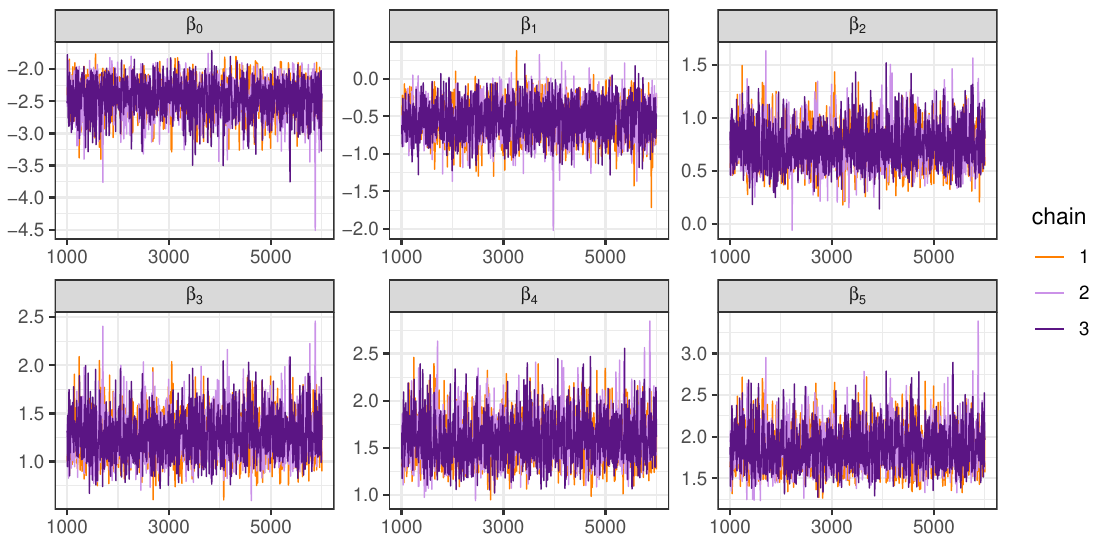}
\captionsetup{labelformat=empty}
\caption{ \textbf{Figure S2c:} Post warm-up MCMCs, using the Bayesian GMM to estimate HR based on the event-free survival in EWING ($R2_{loc}$) trial. $\beta_0$ is the intercept, $\beta_1$ is the parameter of the treatment factor and $(\beta_2, ... \beta_5)$ are for the $K-1$ dummy time points variables.}
\end{figure}

\begin{figure}\centering
\includegraphics[width=15cm]{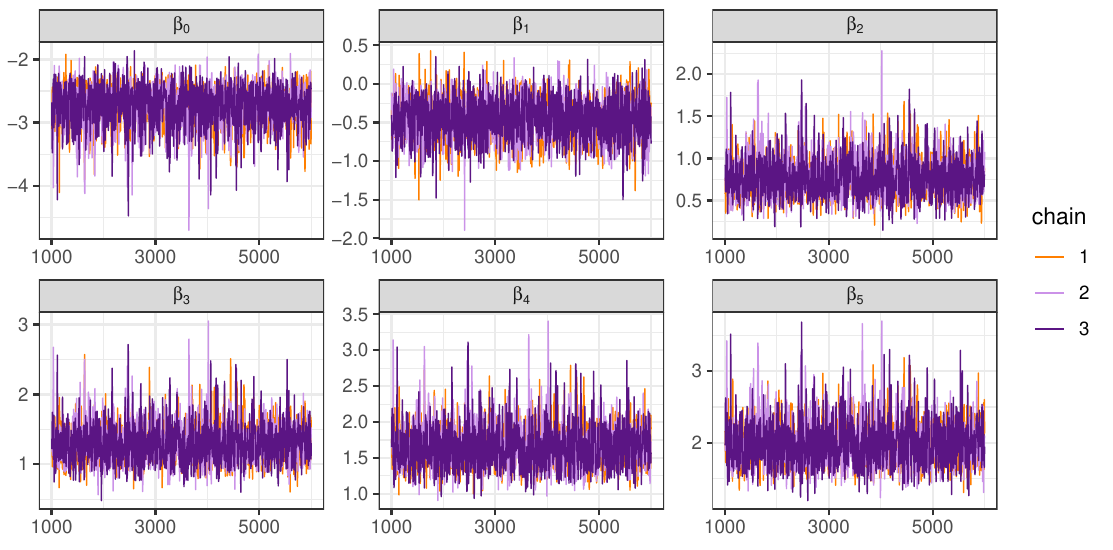}
\captionsetup{labelformat=empty}
\caption{ \textbf{Figure S2d:} Post warm-up MCMCs, using the Bayesian GMM to estimate HR based on the overall survival in EWING ($R2_{loc}$) trial. $\beta_0$ is the intercept, $\beta_1$ is the parameter of the treatment factor and $(\beta_2, ... \beta_5)$ are for the $K-1$ dummy time points variables.}
\end{figure}

\begin{figure}\centering
\includegraphics[width=15cm]{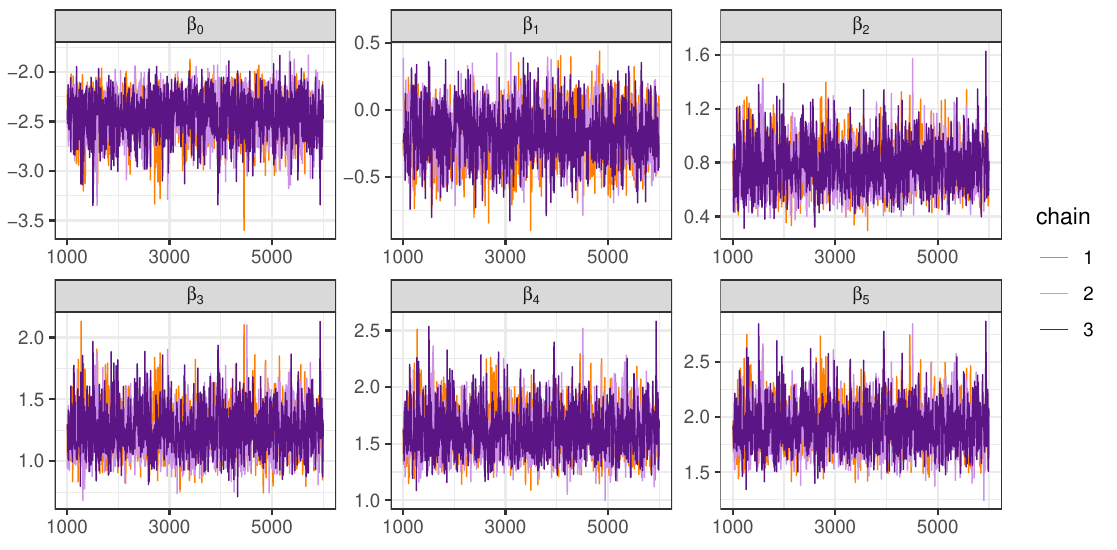}
\captionsetup{labelformat=empty}
\caption{ \textbf{Figure S2e:} Post warm-up MCMCs, using the Bayesian GMM to estimate HR based on the event-free survival in EWING ($R2_{pulm}$) trial. $\beta_0$ is the intercept, $\beta_1$ is the parameter of the treatment factor and $(\beta_2, ... \beta_5)$ are for the $K-1$ dummy time points variables.}
\end{figure}

\begin{figure}\centering
\includegraphics[width=15cm]{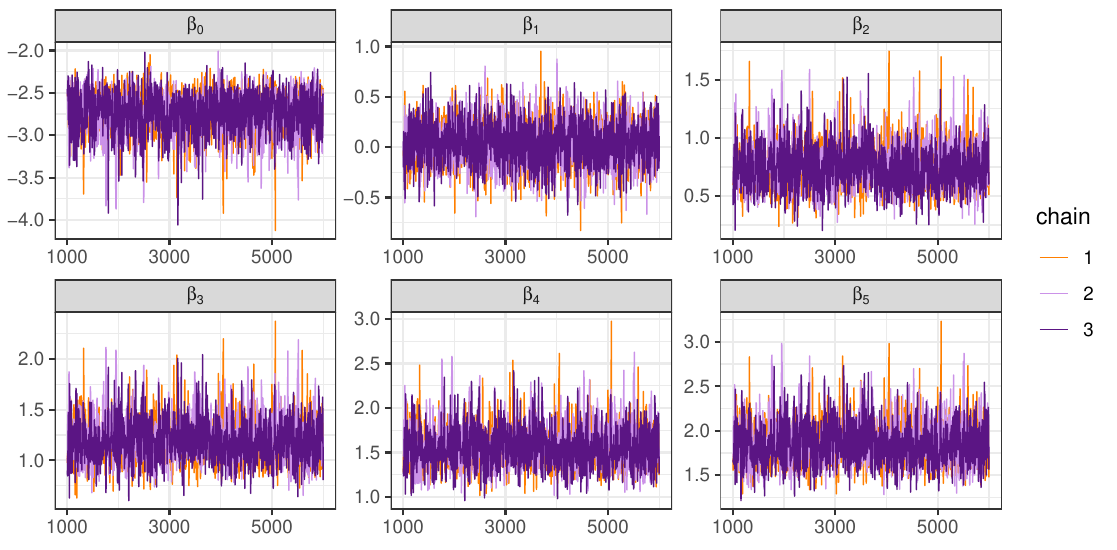}
\captionsetup{labelformat=empty}
\caption{ \textbf{Figure S2f:} Post warm-up MCMCs, using the Bayesian GMM to estimate HR based on the overall survival in EWING ($R2_{pulm}$) trial. $\beta_0$ is the intercept, $\beta_1$ is the parameter of the treatment factor and $(\beta_2, ... \beta_5)$ are for the $K-1$ dummy time points variables.}
\end{figure}

\begin{figure}\centering
\includegraphics[width=17.5cm]{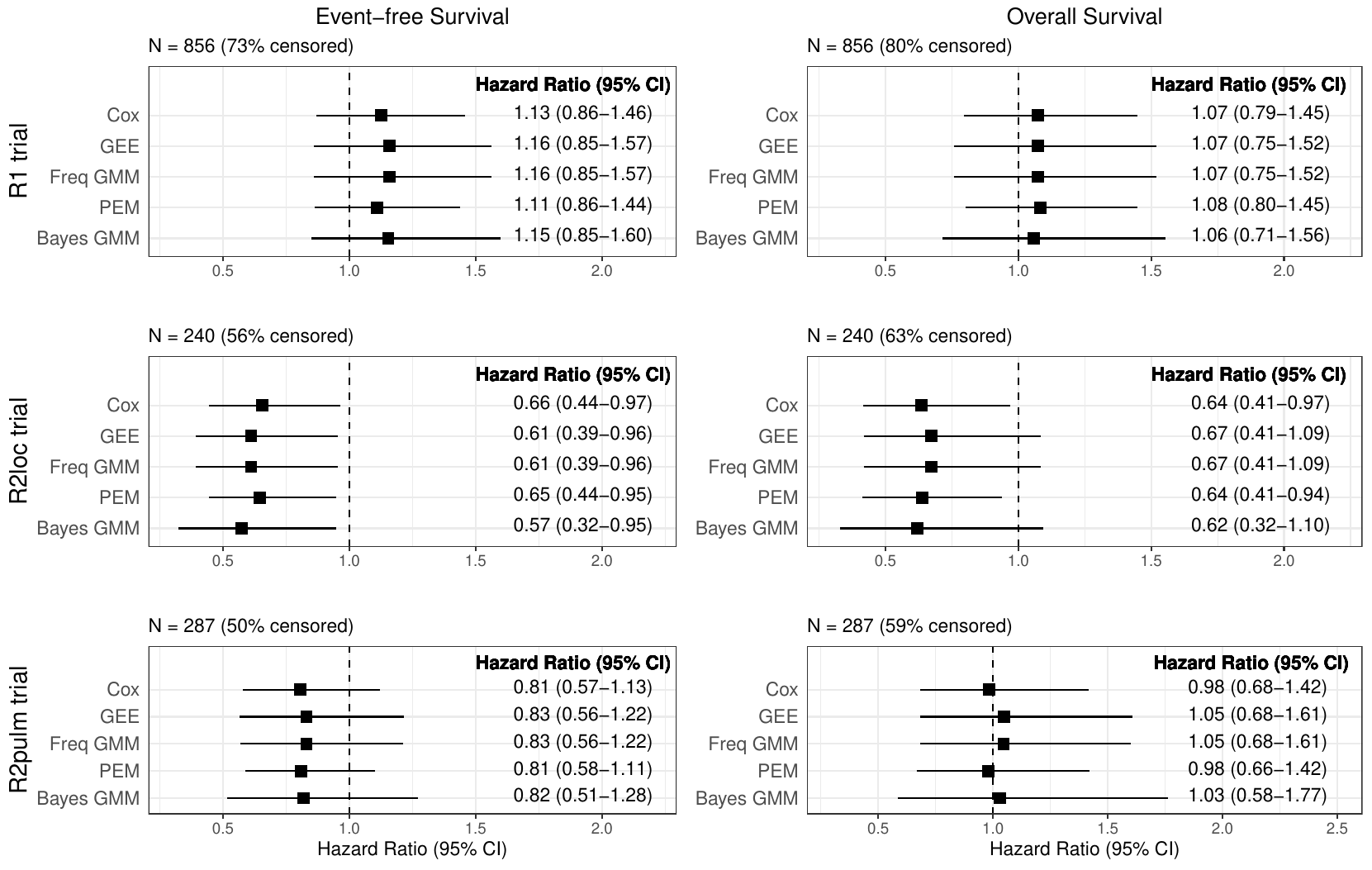}
\captionsetup{labelformat=empty}
\caption{ \textbf{Figure S3:} Hazard ratio estimates (and 95\% confidence intervals) from the Cox proportional hazard, GEE, frequentist (Freq) GMM, piecewise exponential and Bayesian (Bayes) GMM models in the three EWING trials ($R1$, $R2_{loc}$, or $R2_{pulm}$) for event-free survival (left part) and overall survival (right part). These analyses were adjusted on the age variable: binary variable ($<25$, $\geq 25$) years in the $R1$ trial and categorical variable ($<12$, $12-18$, $18-25$, $>25$) years in the $R2_{loc}$, and $R2_{pulm}$ trials. The independent working correlation matrix is used for GEE and GMM approaches. The vertical dashed line represents the null effect.}
\end{figure}

\begin{figure}\centering
\includegraphics[width=12cm]{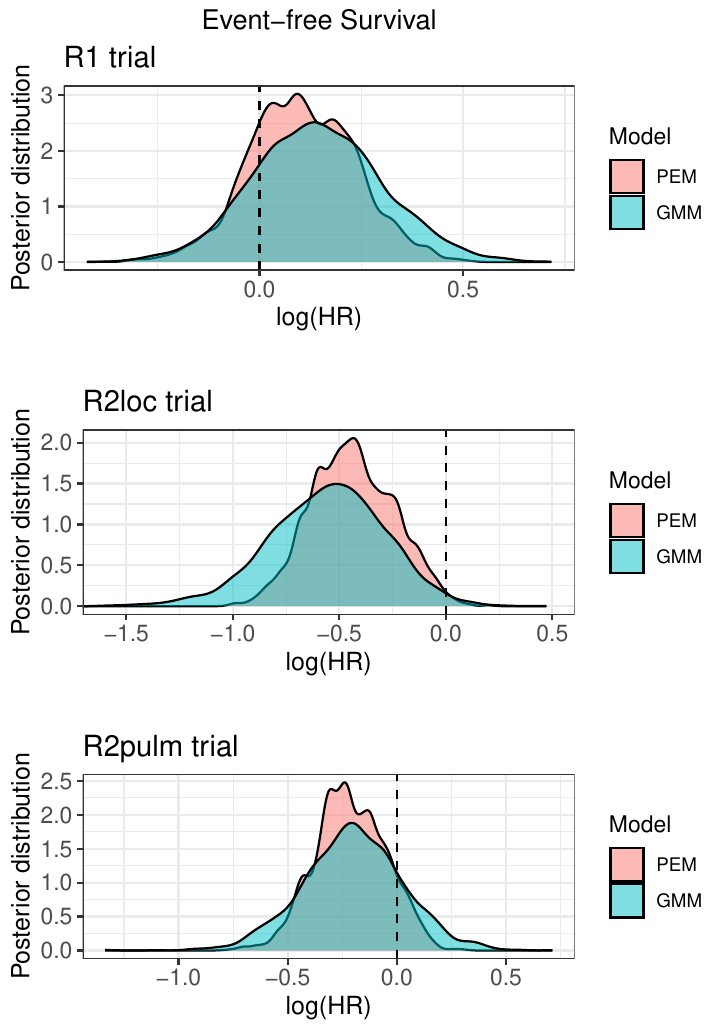}
\captionsetup{labelformat=empty}
\caption{ \textbf{Figure S4:} Posterior distribution of the log hazard ratios estimated with the piecewise exponential model (PEM) and the Bayesian generalized method of moments (GMM) in the three EWING trials ($R1$, $R2_{loc}$, or $R2_{pulm}$) for event-free survival. These analyses were adjusted on the age variable: binary variable ($<25$, $\geq 25$) years in the $R1$ trial and categorical variable ($<12$, $12-18$, $18-25$, $>25$) years in the $R2_{loc}$, and $R2_{pulm}$ trials. The independent working correlation matrix is used for GEE and GMM approaches. The vertical dashed line represents the null effect.}
\end{figure}